\documentclass[twocolumn,secnumarabic,amssymb, nobibnotes, aps, prd]{revtex4-2}

\usepackage{amsmath, amssymb, braket}
\usepackage{xcolor}
\usepackage{bbold}
\usepackage{hyperref}
\usepackage{comment}
\usepackage{mathrsfs}

\usepackage{tikz}  
\usetikzlibrary{arrows,shapes,positioning,shadows,svg.path,backgrounds,fit}

\usepackage{pgfplots}
\usepackage{pgfplotstable}
\pgfplotsset{compat=1.18}
\begin{document}

\title{Frozen and Growing Quantum Work under Noise: Coherence and Correlations as Key Resources}%

\author{Mohammad B. Arjmandi}%
\email[]{mohammad.arjmandi@upol.cz}
\affiliation{Department of Optics, Palack\'y University, 17. listopadu 12, 779 00 Olomouc, Czech Republic}

\begin{abstract}
We explore the decomposition of ergotropy into incoherent and coherent parts for quantum systems exposed to typical Markovian noise channels. The incoherent contribution arises from population inversion in the energy eigenbasis after dephasing, while the coherent contribution reflects the impact of quantum coherence on work extraction. For single-qubit systems, we derive explicit conditions for freezing and even enhancement of the coherent work and establish an analytical upper bound, showing that the coherent contribution cannot exceed one half of the state’s quantum coherence. We then analyze two distinct classes of separable two-qubit states, where both qubits are affected by noise. For Bell-diagonal states—locally completely passive and therefore lacking local coherence—we prove a theorem establishing that the total extractable work under noise equals the average of the geometric quantum and classical correlations. In this case, no enhancement of coherent ergotropy occurs, although specific freezing behavior is identified under certain noise conditions. In contrast, for separable states endowed with local coherence, coherent ergotropy can increase under all investigated noises, even phase flip and depolarizing channels that are typically expected to destroy coherence. Extending to multipartite systems, we find that both the magnitude and the range of noise-induced enhancement scale with the number of qubits, revealing a collective reinforcement of the coherent ergotropy. Finally, we show by means of an explicit example that entanglement does not preclude this enhancement, and that coherent ergotropy can increase under noise even for entangled states. These findings provide novel insights into how noise can be leveraged in quantum systems for energy storage, contrary to conventional perspective, suggesting that noise-assisted enhancement of stored energy can coexist with fast-charging mechanisms typically enabled by entangling operations in quantum batteries.
\end{abstract}

\maketitle

\section{Introduction}
\label{introduction}
Quantum technologies have opened the way for employing microscopic systems as energy storage devices, commonly referred to as quantum batteries~\cite{alicki2013entanglement,campaioli2018quantum,campaioli2024colloquium,bhattacharjee2021quantum}. These systems can surpass their classical counterparts by offering higher efficiency~\cite{dou2022highly,rinaldi2025reliable,ge2023efficient,andolina2019quantum,xu2021enhancing,santos2020stable,shokri2025entanglement,shaghaghi2022micromasers} and, in particular, markedly faster charging rates—an advantage that can, in principle, be counterintuitively amplified with the size of the battery, namely, the number of constituent quantum cells~\cite{campaioli2017enhancing,binder2015quantacell,andolina2025genuine,rossini2020quantum,gyhm2022quantum,ferraro2018high,arjmandi2022performance}. Such intriguing prospects have stimulated extensive research into exploiting distinctively quantum resources— including entanglement~\cite{shokri2025entanglement,shi2022entanglement,kamin2020entanglement,gyhm2024beneficial,imai2023work}, coherence~\cite{shi2022entanglement,mayo2022collective,arjmandi2023localization,lai2024quick,ccakmak2020ergotropy,francica2020quantum}, and many-body interactions~\cite{rossini2019many,guo2024analytically,le2018spin}—aimed at enhancing the performance, functionality, and practical viability of these devices.

Work extraction from quantum batteries has been mostly restricted within the constraint of unitary evolution, which preserves entropy and precludes dissipation. Under these conditions, the maximum usable energy is quantified by the ergotropy~\cite{allahverdyan2004maximal}. A quantum state is said to be passive if, under all possible unitary operations, its energy cannot be lowered—implying that it contains no accessible work in this idealized setting~\cite{allahverdyan2004maximal,pusz1978passive}.

Despite the promises, a major challenge remains: energy loss to the environment, typically known as self-discharging~\cite{santos2021quantum,song2025self} or anomalous discharging~\cite{medina2025anomalous}. Such processes drastically limit the time over which the stored energy can be usefully extracted. Several strategies have been proposed to combat this effect, such as exploiting quantum coherence combined with locally applied disordered magnetic field to hinder energy leakage~\cite{arjmandi2022enhancing}, or employing dark states that remain invariant under environmental coupling~\cite{liu2019loss,quach2020using}, thereby preserving stored energy. Other approaches include the development of a feasible scheme based on nitrogen–vacancy (NV) center in diamond in which the hyperfine interaction between electronic and nuclear spins provides additional coherent control to prevent the stored energy from dissipation for longer durations~\cite{song2025self}. Similarly, non-Markovian effects also offer a potential advantage by enabling energy backflow from the environment, thereby reducing the rate of unwanted discharging~\cite{kamin2020non,xu2024inhibiting}.

While the adverse effects of noise on quantum batteries are partially addressed, the detailed dynamics of work extraction under specific noise models remain an active area of research. It has been demonstrated that dephasing can facilitate faster charging of quantum batteries within certain parameter regimes~\cite{shastri2025dephasing}. Non-local operations are also introduced to partially recover the work content of the quantum battery, after being affected by noise~\cite{tirone2023work}. Furthermore, presence of quantum coherence appears to be beneficial for sustaining work capacitance—defined as the maximal average work extractable per system from an ensemble of quantum batteries under self-discharging~\cite{tirone2025quantum}, although no increase in any measure of extractable energy under noise is observed.

On the other hand, the extractable energy, quantified either by ergotropy or free energy difference, can be decoposed into coherent and incoherent contribution~\cite{shi2022entanglement,francica2020quantum}. The latter corresponds to the ergotropy of the state after full dephasing in the energy eigenbasis, representing work due solely to population inversion. Then, the coherent contribution is the remaining part, i.e., the difference between the total and incoherent ergotropy. The positive interplay between quantum coherence and coherent ergotropy is experimentally verified in a noiseless setting, using a single-qubit platform based on a nitrogen–vacancy (NV) center in diamond~\cite{niu2024experimental}. Although this coherent part reflects the influence of coherence in work extraction, it is not a coherence monotone as it may increase under incoherent coherence-preserving operations~\cite{francica2020quantum}. 

In~\cite{shi2022entanglement}, the authors examined a noiseless battery–charger scenario and found that the incoherent contribution to extractable work is negatively correlated with both battery–charger entanglement and the battery’s coherence, whereas the coherent contribution increases with the battery’s internal coherence. Their study, however, is limited to closed, unitary dynamics and does not consider the role of environmental noises.

In this work, we take a step further by splitting the total ergotropy into its incoherent and coherent contributions for quantum systems subjected to common Markovian noise channels: bit-flip, bit-phase-flip, phase-flip, depolarizing, and amplitude damping~\cite{nielsen2010quantum,mcmahon2008quantum,preskill1998lecture}. These channels describe memory-less interactions between a quantum system and its environment. Flip errors—such as bit flip, phase flip, and bit-phase flip—respectively disturb the occupation probabilities, the relative phases, or both, thereby corrupting the encoded quantum information. The depolarizing channel, arising from coupling to a highly chaotic bath, erases all information and drives the system toward the maximally mixed state. On the other hand, amplitude damping embodies energy dissipation, modeling processes such as spontaneous emission, where the system irreversibly loses excitations to its surroundings.

For single-qubit systems, we identify specific conditions—expressed in terms of the system’s parameters and noise strengths—under which the work component can remain frozen or, remarkably, increase with noise. The latter behavior does not occur for the incoherent part and is absent in the depolarizing channel, which drives the system to a maximally mixed state which is completely passive, incapable of delivering work through global unitaries on an ensemble of identical copies. In addition, we prove that the coherent ergotropy is upper bounded by one half of quantum coherence. We further generalize our study to non interacting two-qubit systems {by considering two classes of separable states. The first one} is described in a Bell-diagonal form, whose available ergotropy entirely arises from its global features, since its local marginals are passive. Therefore, for these states, we prove a theorem linking total extractable work to the average of geometric quantum and classical correlations, when both subsystems are affected by noise. This theorem, however, does not hold for the amplitude damping channel, which breaks the Bell-diagonal form due to its non unital nature. For these two-qubit systems, while no growing behavior emerges, coherent and incoherent work can remain frozen or retain residual nonzero values at large noise strengths, depending on the channel.

We next consider a distinct class of separable states featuring local coherence, for which the coherent component of ergotropy exhibits substantial enhancement across a broad range of noise strengths for all channels, even under phase flip and depolarizing channel, given an appropriate choice of the system’s energy basis.

Finally, using the same class of states with local coherence, we extend our analysis to multipartite systems and demonstrate that both the maximum  and area of enhancement over noise strength can scale with system size, i.e., the number of embedded qubits.

Furthermore, we provide an explicit example showing that such noise-assisted enhancement is not confined to separable states, but can also be observed for entangled ones.

Our results uncover an unexpected and fundamentally interesting phenomenon: in specific scenarios, noise can act constructively, either enhancing the available work or leaving it unaffected. This stands in sharp contrast to the widely held view that noise solely diminishes quantum thermodynamic performance, highlighting rich and nontrivial dynamics at the intersection of coherence, correlations, and dissipation.

The remainder of the paper is organized as follows. In the next section, we introduces the ergotropy and its decomposition into coherent and incoherent parts. Sec.~\ref{markovian_channels} presents the noise channels within the Kraus operator formalism. In Sec.~\ref{singl_two_qubit}, frozen and noise-assisted growth conditions for work contents are evaluated for single and two-qubit cases. Sec.~\ref{multi} generalize the results to multipartite systems. Finally, we conclude the paper in Sec.~\ref{conc}.

\section{Ergotropy and its components}
\label{erg_contribution}
A quantum state is said to be passive with respect to a relevant Hamiltonian if its energy cannot be lowered by unitary operations. Decomposing the density matrix $\rho=\sum_{i}\lambda_{i}\ket{\lambda_{i}}\bra{\lambda_{i}}$ and Hamiltonian $H=\sum_{i}\epsilon_{i}\ket{\epsilon_{i}}\bra{\epsilon_{i}}$ into their eigenstates with ordering conditions $\lambda_{i}\geq \lambda_{i+1}$ and $\epsilon_{i}\leq \epsilon_{i+1}$, the corresponding passive state $\pi_{\rho}$ is given by $\pi_{\rho}=\sum_{i}\lambda_{i}\ket{\epsilon_{i}}\bra{\epsilon_{i}}$, which lacks coherence and population inversion. We introduce \textit{ergotropic map} $\Phi$ as a unitary and hence, completely positive and trace-preserving operation such that $\Phi(\rho)=\pi_{\rho}$. Then, ergotropy as the maximum extractable energy from $\rho$ with respect to $H$ under unitary operation is the energy difference between $\rho$ and its converted form under the ergotropic map $\Phi$, i.e. the passive state $\pi_{\rho}$
\begin{align}
\mathcal W(\rho)={\rm Tr} \left[H (\rho - \pi_{\rho}) \right].\label{erg_def}
\end{align}
Now, consider \textit{incoherent ergotropic map} $\Phi^{\rm I}$ to be another unitary operation which can eliminate the population inversion of $\rho$ without altering its coherence. The extracted energy by $\Phi^{\rm I}$ is called incoherent ergotropy which is equivalent to the ergotropy of the dephased state $\zeta(\rho)$ by energy eigenbasis $\zeta(\rho)=\sum_{i} \bra{\epsilon_{i}} \rho \ket{\epsilon_{i}} \ket{\epsilon_{i}} \bra{\epsilon_{i}}$
\begin{align}
\mathcal W^{\rm I}(\rho)=\mathcal W(\zeta(\rho)).\label{incerg_def}
\end{align}
Finally, the difference between total and incoherent ergotropy is called coherent ergotropy~\cite{shi2022entanglement,francica2020quantum}
\begin{align}
\mathcal W^{\rm C}(\rho)=\mathcal W(\rho)-\mathcal W^{\rm I}(\rho)={\rm Tr} \left[H (\pi_{\zeta(\rho)} - \pi_{\rho}) \right],\label{coerg_def}
\end{align}
where in the last equality, we have used the fact that $\rho$ and its dephased form $\zeta(\rho)$ share the same diagonal elements and hence internal energy.
\section{Markovian channels}
\label{markovian_channels}
In this section, we briefly introduce some well-known types of Markovian noise that commonly affect quantum systems. The evolution of a quantum system under such noises can be effectively described using the Kraus operator formalism~\cite{kraus1971general,kraus1983states}. Given an initial density matrix $\rho$, its evolution under Markovian quantum channel $\Lambda$ is given by~\cite{nielsen2010quantum}
\begin{align}
\Lambda(\rho)=\sum_{j} K_{j} \rho K_{j}^{\dagger},\label{markovian_kraus}
\end{align}
where ${K_{j}}$ are the Kraus operators satisfying the completeness relation  $\sum_{j} K_{j}^{\dagger} K_{j}=\mathrm{I}$ where $\mathrm{I}$ is the identity operator. Using the basis $\ket{\rm g}=\left( \begin{matrix} 1 \\ 0 \end{matrix} \right)$ and $\ket{\rm e}=\left( \begin{matrix} 0 \\ 1 \end{matrix} \right)$, we introduce several channels, characterized by the noise parameter $0\leq q \leq 1$. Bit flip, bit-phase flip and phase flipe are respectively defined by a pair of Kraus operators as $(K_{0}^{\rm f},K_{1}^{\rm f})$, $(K_{0}^{\rm f},K_{2}^{\rm f})$, and $(K_{0}^{\rm f},K_{3}^{\rm f})$ where
\begin{align}
K_{0}^{\rm f}=\sqrt{1-q/2}\mathrm{I}, \quad K_{i}^{\rm f}=\sqrt{q/2}\sigma_{i},\label{flip_noise}
\end{align}
in which $\sigma_{i}$ ($i=1, 2, 3$) respectively correspond to the standard Pauli matrices $\sigma_{\rm x}$, $\sigma_{\rm y}$, and $\sigma_{\rm z}$~\cite{farina2019charger,orszag2008quantum}. The depolarizing channel is similarly defined by four Kraus operators
\begin{align}
K_{0}^{\rm d}=\sqrt{1-q/4}\mathrm{I}, \quad K_{i}^{\rm d}=\sqrt{q/4}\sigma_{i}.\label{depolarizing}
\end{align}
Finally, the amplitude damping and phase damping channels are also defined by a pair of $(K_{0}^{\rm a},K_{1}^{\rm a})$ and $(K_{0}^{\rm p},K_{1}^{\rm p})$, respectively given by
\begin{align}
K_{0}^{\rm a}=\ket{\rm g}\bra{\rm g}+\sqrt{1-q}\ket{\rm e}\bra{\rm e}, \quad K_{1}^{\rm a}=\sqrt{q}\ket{\rm g}\bra{\rm e},\label{amplitude damping}
\end{align}
and
\begin{align}
K_{0}^{\rm p}=\ket{\rm g}\bra{\rm g}+\sqrt{1-q}\ket{\rm e}\bra{\rm e}, \quad K_{1}^{\rm p}=\sqrt{q}\ket{\rm e}\bra{\rm e}.\label{amplitude damping}
\end{align}
Since phase damping and the phase-flip channel lead to essentially similar dephasing dynamics, we only concentrate on the phase-flip channel in the present work.
\section{Frozen and growing extractable work}
\label{singl_two_qubit}
In this section, we examine how the previously introduced channels influence coherent and incoherent ergotropy, uncovering specific conditions that lead the ergotropy component to freeze or even increase under noise in some scenarios. We begin by analyzing a single-qubit system and then expand our investigation to a two-qubit system.
\subsection*{A: single qubit system}
\begin{figure}[h!]
\begin{center}
\includegraphics[width=1\columnwidth]{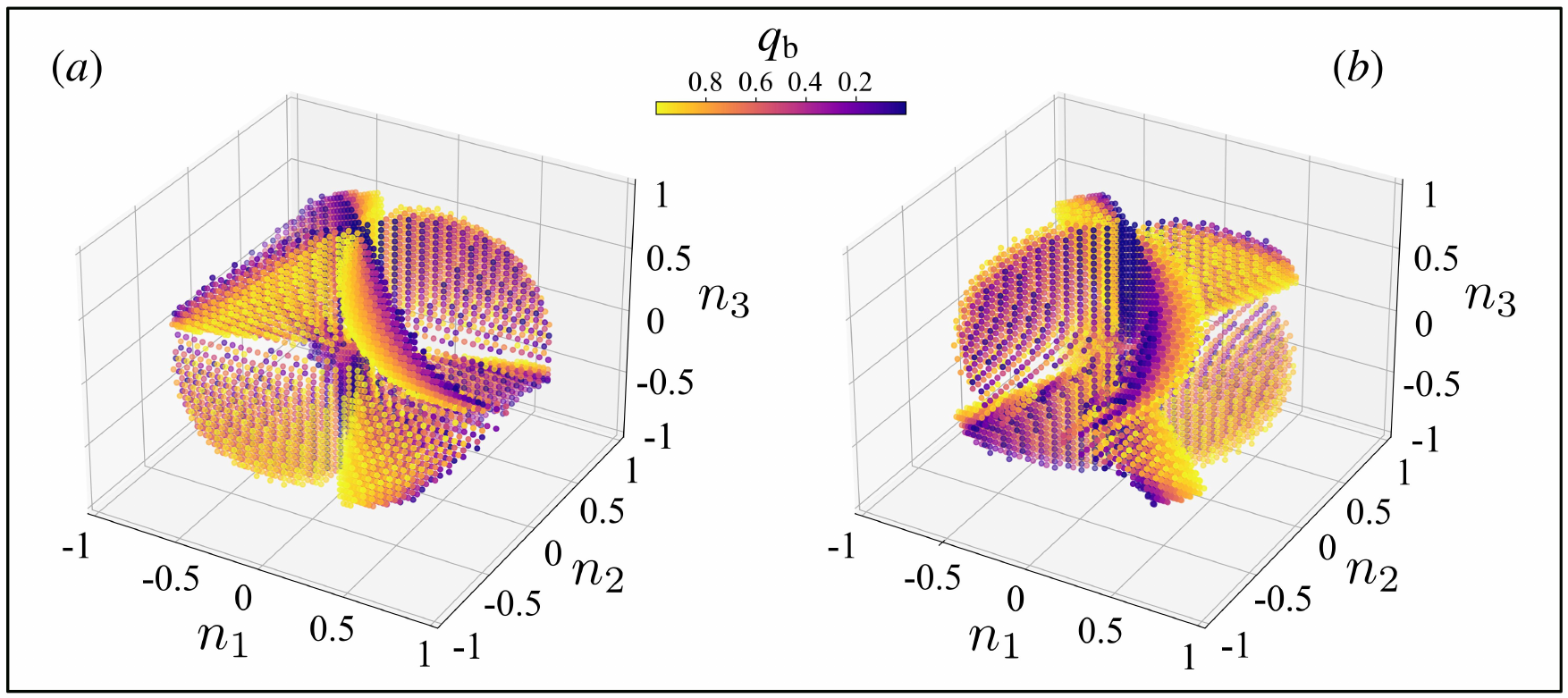}
\end{center}
\caption{Three-dimensional plot of the noise strength bound resulting in coherent ergotropy enhancement for (a) bit flip Eq.~\eqref{q_frozen_growing_bit_flip_single} and (b) bit-phase flip Eq.~\eqref{q_frozen_growing_bit_phase_flip_single}.}
\label{3d_q_condition_bit_flip_bit_phase_flip}
\end{figure}

\begin{figure}[h!]
\begin{center}
\includegraphics[width=1\columnwidth]{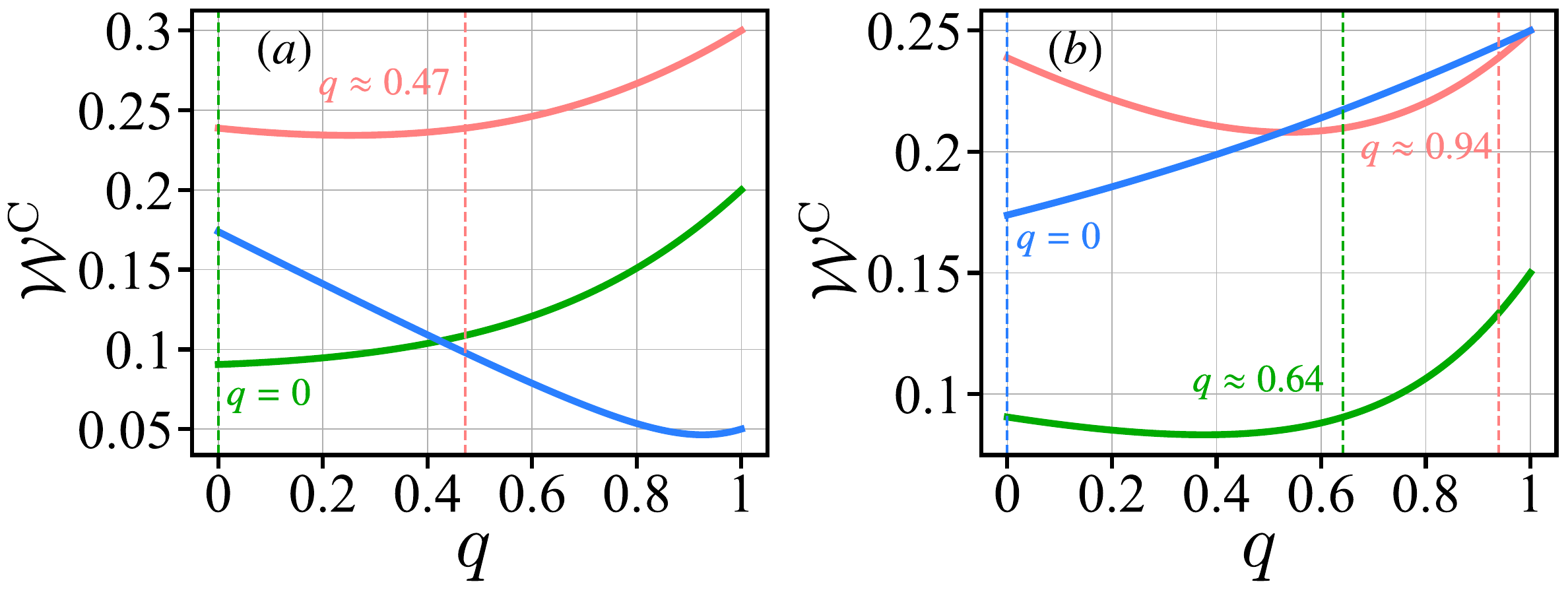}
\end{center}
\caption{Coherent ergotropy of single qubit system under (a) bit flip Eq.~\eqref{coerg_bit_flip_single} and (b) bit-phase flip. The vertical dashed lines pinpoint the noise strength for which the enhancement of coherent ergotropy with respect to its noiseless value occurs, i.e. $\mathcal{W}^{\rm C}(q) \geq \mathcal{W}^{\rm C}(0)$. The Bloch vector parameters ($n_{1}$, $n_{2}$, $n_{3}$), respectively for salmon, green and blue curves are ($0.6$, $0.5$, $0.4$), ($0.4$, $0.3$, $0.6$), and ($0.1$, $0.5$, $0.2$).}
\label{bit_flip_bit_phase_flip_single_qubit}
\end{figure}

Consider the generic Bloch representation of a single qubit as~\cite{barnett2009quantum}
\begin{align}
\rho=\frac{1}{2}(\mathrm{I} + \mathbf{n} \cdot \boldsymbol{\sigma}),\label{bloch-single-qubit}
\end{align}
where $\mathbf{n}=\{ n_{1},n_{2},n_{3} \}$ is the Bloch vector satisfying $\lVert \mathbf{n} \rVert = \sqrt{n_{1}^{2}+n_{2}^{2}+n_{3}^{2}} \leq 1$ and $\boldsymbol{\sigma}=\{ \sigma_{1},\sigma_{2},\sigma_{3} \}$ denotes the vector of Pauli operators. Without loss of generality, we assume the energy of system is described by $H_{0} = B \ket{\rm e}\bra{\rm e}$ where $B$ is the energy gap between the ground and excited levels. Thorough the rest of paper, we express the extractable work in the unit of $B$ so we can simply assume that $B=1$. Using Eq.~\eqref{erg_def}, the total ergotropy of the above state can be obtained as
\begin{align}
\mathcal W(\rho)=\frac{1}{2}(\lVert \mathbf{n} \rVert - n_{3}).\label{total_erg_single_qubit}
\end{align}
As $\rho$ is already written in the energy (computational) basis, one just needs to wash out the off-diagonal elements to obtain the dephased form of $\rho$ with energy eigenstate, i.e., $\zeta(\rho)=\frac{1+n_{3}}{2} \ket{\rm g}\bra{\rm g}+\frac{1-n_{3}}{2} \ket{\rm e}\bra{\rm e}$ which provides useful extractable energy due to population inversion if $n_{3}<0$. It then follows that the incoherent ergotropy is $\mathcal W^{\rm I}(\rho)=\lvert n_{3} \rvert$. Finally, the coherent ergotropy simply reads
\begin{align}
\mathcal W^{\rm C}(\rho)=\frac{1}{2}(\lVert \mathbf{n} \rVert - \lvert n_{3} \rvert).\label{coerg_single_qubit}
\end{align}
We note that when $n_{3}>0$ then $\mathcal W(\rho)=\mathcal W^{\rm C}(\rho)$ as energy cannot be extracted by incoherent ergotropic map $\Phi^{\rm I}$ due to the lack of population inversion in $\zeta(\rho)$. In what follows, we separately investigate the effect of each channel on Eq.~\eqref{bloch-single-qubit} and its extractable energy content.\\
\\
\textbf{Bit flip}. Using the Kraus operator formalism defined by Eq.~\eqref{markovian_kraus}, the bit flip noise transforms the Bloch vector into $\mathbf{n}^{\rm bf}=\{ n_{1},n_{2} (1-q),n_{3} (1-q) \}$, such that the first component, which encodes coherence along x-axis, does not evolve. Trivially, the state and its available energy content remain invariant under the bit flip provided $n_{2}=n_{3}=0$. One can immediately obtain the incoherent ergotropy as
\begin{align}
\mathcal W^{\rm I}(\rho^{\rm bf}) &= 
\begin{cases}
|n_3|(1 - q) &  n_3 < 0 \\[6pt]
\quad 0, & \text{otherwise}
\end{cases}
\label{incerg_bit_flip_single}
\end{align}
which is consistently decreasing with the noise strength $q$ and vanishes at $q=1$ as in this limit the dephased state $\zeta(\rho^{\rm bf})$ becomes maximally mixed and hence, passive (indeed completely passive as for single-qubit system, the maximally mixed state is nothing but thermal). The energy of passive states $\pi_{\rho^{\rm bf}}$ and $\pi_{\zeta(\rho^{\rm bf})}$ are given by ${\rm Tr}\left[ H_{0} \pi_{\rho^{\rm bf}} \right]=\frac{1}{2}(1 - \lVert \mathbf{n}^{\rm bf} \rVert)$ and ${\rm Tr} \left[ H_{0} \pi_{\zeta(\rho^{\rm bf})} \right]=\frac{1}{2}(1 - \lvert n_{3} \rvert (1-q))$, respectively. Using Eq.~\eqref{coerg_def}, these result in the following relation for the coherent ergotropy
\begin{align}
\mathcal W^{\rm C}(\rho^{\rm bf})=\frac{1}{2}(\lVert \mathbf{n}^{\rm bf} \rVert - \lvert n_{3} \rvert (1-q)).\label{coerg_bit_flip_single}
\end{align}
It is straightforward to verify that $\mathcal W^{\rm C}(\rho^{\rm bf})\big|_{q=0} = \mathcal W^{\rm C}(\rho)$ where $\mathcal W^{\rm C}(\rho)$ is the coherent ergotropy of the initial state $\rho$ given by Eq.~\eqref{coerg_single_qubit}. A trivial freezing condition for coherent ergotropy is $n_{2}=n_{3}=0$ which yields $\mathcal W^{\rm C}(\rho^{\rm bf})=\frac{n_{1}}{2}$, being noise-independent since in this case the initial state is invariant under bit flip channel, as previously mentioned. On the other hand, it can be shown that the necessary condition to have frozen or increasing ergotropy under bit flip noise, i.e., $\mathcal W^{\rm C}(\rho^{\rm bf}) \geq \mathcal W^{\rm C}(\rho)$ is that both $n_{1}$ and $n_{3}$ are non zero, because if $n_{1}=0$, then $\mathcal W^{\rm C}(\rho^{\rm bf})\big|_{n_{1}=0}=\frac{1-q}{2}(\sqrt{n_{2}^{2}+n_{3}^{2}} - \lvert n_{3} \rvert) \leq \mathcal W^{\rm C}(\rho)\big|_{n_{1}=0} = \frac{1}{2}(\sqrt{n_{2}^{2}+n_{3}^{2}} - \lvert n_{3} \rvert)$ and similarly, if $n_{3}=0$ then $\mathcal W^{\rm C}(\rho^{\rm bf})\big|_{n_{3}=0}=\frac{1}{2}(\sqrt{n_{1}^{2}+(1-q)^{2}n_{2}^{2}}) \leq \mathcal W^{\rm C}(\rho)\big|_{n_{3}=0} = \frac{1}{2}(\sqrt{n_{1}^{2}+n_{2}^{2}})$. Indeed, having frozen or growing coherent ergotropy under bit flip sets a condition on the noise strength as
\begin{align}
q \geq q_{\rm b}^{\rm bf}=\frac{2(n_{2}^{2}+n_{3}^{2}-\lvert n_{3} \rvert \lVert \mathbf{n} \rVert)}{n_{2}^{2}},\label{q_frozen_growing_bit_flip_single}
\end{align}
up to selecting appropriate Bloch vector components resulting in valid degrees of noise strength such that $q \in \left[0,1 \right]$. In Fig.~\ref{3d_q_condition_bit_flip_bit_phase_flip}a the bound $q_{\rm b}^{\rm bf}$ is depicted, highlighting the region in which the coherent ergotropy is preserved or enhanced under the bit flip channel. In addition, Fig.~\ref{bit_flip_bit_phase_flip_single_qubit}a illustrates how coherent ergotropy varies with bit flip noise strength $q$, with outcomes depending on the structure of the Bloch vector. 
According to the condition Eq.~\eqref{q_frozen_growing_bit_flip_single}, when $n_{2}^{2}+n_{3}^{2}<\lvert n_{3} \rvert \lVert \mathbf{n} \rVert)$ then $q_{\rm b}^{\rm bf}$ is negative, implying that all physically valid noise strengths ($0 \leq q \leq 1$) result in growth of the coherent ergotropy such that $\mathcal W^{\rm C}(\rho^{\rm bf})>\mathcal W^{\rm C}(\rho)$, as seen by the green curve for which $q_{\rm b}^{\rm bf} \approx -0.41$. For the red curve, however, $q_{\rm b}^{\rm bf} \approx 0.47$ (dashed vertical line), indicating that coherent ergotropy grows only when the noise strength surpasses this value. Finally, for the blue curve, we have $q_{\rm b}^{\rm bf} \approx 1.4$, which exceeds the maximum allowed value of $q$, meaning that the always $\mathcal W^{\rm C}(\rho^{\rm bf})<\mathcal W^{\rm C}(\rho)$. These results are particularly intriguing, as they reveal that strong enough bit-flip noise can enhance coherent ergotropy—contrary to the usual expectation that noise suppresses extractable work.\\

Moreover, the quantum coherence of $\rho^{\rm bf}$ can be quantified using the $l_1$-norm measure $\mathcal{C}(\rho) = \sum_{i \neq j}{\lvert \rho_{ij} \rvert}$~\cite{baumgratz2014quantifying,streltsov2017colloquium} as
\begin{equation}
\mathcal{C}(\rho^{\rm bf}) = \sqrt{n_1^2 + n_2^2 (1 - q)^2}.\label{coherence_bf}
\end{equation}
It can be shown that the coherent contribution to the ergotropy is upper bounded by the state’s coherence,
\begin{equation}
\mathcal{W}^{\rm C} \le \frac{1}{2}\mathcal{C},\label{upper_bound_coherence_bf}
\end{equation}
as detailed in the Appendix~\ref{app_A}. In addition, it can be readily verified that when $q=1$, then $\mathcal{W}^{\rm C} = \frac{1}{2}\mathcal{C}$, implying that the ultimate value of the coherent work under noise is directly proportional to the residual quantum coherence. This final value, however, does not necessarily coincide with the maximum of $\mathcal{W}^{\rm C}$, since both coherence and work may decay monotonically with increasing noise while asymptotically approaching the same value. Another useful relation arises for states with $n_3 = 0$, where $\mathcal{W}^{\rm C}(q) = \tfrac{1}{2}\mathcal{C}(q)$ holds for all $q$.

\textbf{Bit-phase flip}. In the presence of the bit-phase flip noise, the Bloch vector changes as $\mathbf{n}^{\rm bpf}=\{ n_{1} (1-q),n_{2},n_{3} (1-q) \}$. This mirrors the bit flip case, except the unaffected coherence component is now $n_{2}$ instead of $n_{1}$. Accordingly, all previous results apply to the bit-phase flip noise by replacing these two Bloch vector components. Therefore, the freezing and increasing condition for coherent ergotropy is
\begin{align}
q \geq q_{\rm b}^{\rm bpf}=\frac{2(n_{1}^{2}+n_{3}^{2}-\lvert n_{3} \rvert \lVert \mathbf{n} \rVert)}{n_{1}^{2}}.\label{q_frozen_growing_bit_phase_flip_single}
\end{align}
Figs.~\ref{3d_q_condition_bit_flip_bit_phase_flip}b and~\ref{bit_flip_bit_phase_flip_single_qubit}b summarize the the results obtained under bit-phase flip channel, for which similar discussion to the bit flip noise holds.\\
In analogy with the bit-flip case, one can verify that the upper bound $\mathcal{W}^{\rm C} \leq \frac{1}{2} \mathcal{C}$ also holds for the bit-phase-flip channel by simply interchanging $n_{1}$ and $n_{2}$ in the corresponding expressions.
\\
\textbf{Phase flip}. By its definition, the phase flip channel only affects the off-diagonal element of the density matrix such that the Bloch vector becomes $\mathbf{n}^{\rm pf}=\{ n_{1} (1-q),n_{2} (1-q),n_{3} \}$. As expected, the incoherent part of ergotropy—which reflects the stored energy in population inversion—remains unaffected and hence $W^{\rm I}(\rho^{\rm pf})=W^{\rm I}(\rho)=\lvert n_{3} \rvert$ provided $n_{3}<0$. However, this is not the case for the coherent ergotropy
\begin{align}
\mathcal W^{\rm C}(\rho^{\rm pf})=\frac{1}{2}(\lVert \mathbf{n}^{\rm pf} \rVert - \lvert n_{3} \rvert),\label{coerg_phase_flip_single}
\end{align}
which is clearly decreasing and as $q \to 1$, then $\mathcal W^{\rm C}(\rho^{\rm pf}) \to 0$. Moreover, imposing the frozen and growing condition to the noise strength yields $q>2$ which is not valid as $q \in [0,1]$. Quantum coherence under phase flip is obtained as 
\begin{equation}
\mathcal{C}(\rho^{\rm pf}) = \sqrt{(n_1^2 + n_2^2) (1 - q)^2},\label{coherence_pf}
\end{equation}
indicating a monotonic loss of coherence with increasing noise strength. Therefore, phase-flip noise inevitably suppresses the coherent contribution to ergotropy by eliminating quantum coherence. Nevertheless, the upper-bound relation $\mathcal{W}^{\rm C} \leq \frac{1}{2} \ \mathcal{C}$ still holds, with equality achieved only when both quantities vanish at $q=1$.

The situation can change qualitatively, if we choose an energy basis such that the Hamiltonian does not commute with the channel operators. In this case, the channel can modify the populations in the energy basis—in contrast to the computational basis where phase-flip leaves them unchanged—and such population reshuffling can potentially increase the coherent part of ergotropy. This non-commutativity is therefore a necessary, though not sufficient, condition for observing an enhancement in coherent work.  

For instance, taking the Hamiltonian $H_{0} = \sigma_x$, whose energy eigenbasis corresponds to the $\rm x$-basis, the coherent ergotropy then becomes
\begin{equation}
\mathcal{W}^{\rm C}_{\rm x}(\rho^{\rm pf}) = \lVert \mathbf{n}^{\rm pf} \rVert - \lvert n_{1} \rvert (1-q),\label{co_erg_pf_x_basis}
\end{equation}
which depending on the parameters, the second term can decay faster than the first one, causing $\mathcal{W}^{\rm C}_{\rm x}(\rho^{\rm pf})$ to increase with the noise strength. The condition for freezing or enhancement of the coherent ergotropy can be obtained as $q \geq q_{\rm b}^{\rm pf}$ where $q_{\rm b}^{\rm pf}$ is given by Eq.~\eqref{q_frozen_growing_bit_flip_single}, by replacing $n_3$ with $n_1$. For brevity, we do not visualize this bound, as it is qualitatively similar to the cases of bit flip and bit-phase flip noises. The quantum coherence in the energy basis also reads
\begin{equation}
\mathcal{C}_{\rm x}(\rho^{\rm pf}) = \sqrt{n_{2}^{2}(1-q)^2 + n_3^2},\label{coherence_pf_x_basis}
\end{equation}
which, although decreasing with noise, retains a non-zero residual value in the large-noise limit $q \to 1$, in contrast to the coherence in the computational basis (Eq.~\eqref{coherence_pf}).

\textbf{Depolarizing channel}. By using the Kraus operators Eq.~\eqref{depolarizing}, the Bloch vector of the system under depolarizing noise evolves as $\mathbf{n}^{\rm dc}=\{ n_{1} (1-q),n_{2} (1-q),n_{3} (1-q) \}$ whose all components are decaying with $q$ leading $\rho$ to a maximally mixed state. In fact, under depolarizing channel, the system loses all of its information and becomes maximally mixed with probability $q$ or remain intact with probability $(1-q)$. Thus, both coherent and incoherent ergotropy decrease with depolarizing noise strength and vanish when $q \to 1$. This result is basis-independent, unlike the case of phase flip channel, as the depolarizing channel contracts the Bloch sphere isotropically in all directions, leading to a uniform degradation of the extractable work.\\

\begin{figure}[h!]
\begin{center}
\includegraphics[width=1\columnwidth]{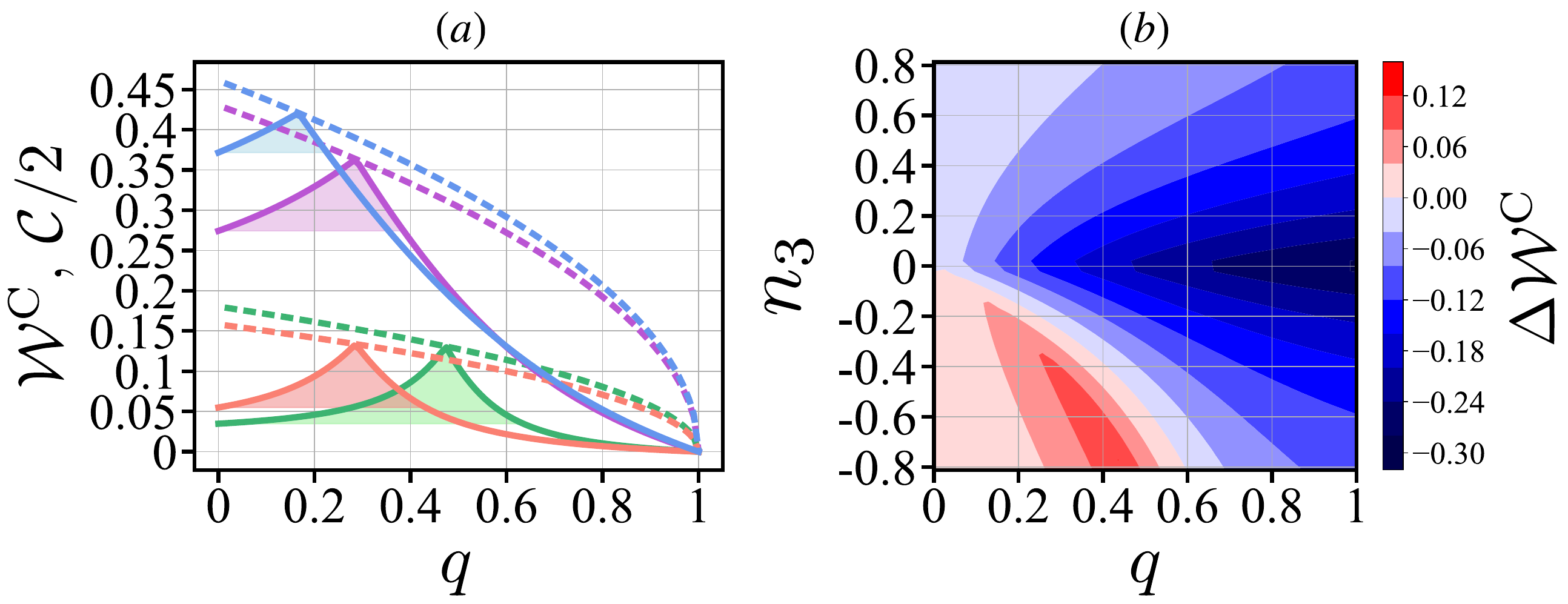}
\end{center}
\caption{(a) Coherent ergotropy (solid) and quantum coherence (dashed) of single qubit system under amplitude damping channel, given by Eq.~\eqref{coerg_amplitude_damping_single}. The shaded areas show the regions for which the coherent ergotropy grows beyond its noise-free value. The Bloch vector parameters ($n_{1}$, $n_{2}$, $n_{3}$) are (0.2, 0.3, -0.9), (0.7, 0.5, -0.4), (0.1, 0.3, -0.4), and (0.6, 0.7, -0.2) for green, purple, salmon and blue curves, respectively. (b) The difference in coherent ergotropy, with and without amplitude damping. The areas for which $\Delta \mathcal W^{\rm C}>0$ show the enhancement of coherent ergotropy of single qubit system subjected to amplitude damping channel. For this panel, we fix $n_{1}=0.5$ and $n_{2}=0.3$.}
\label{plot_amplitude_damping_single_qubit}
\end{figure}

\textbf{Amplitude damping}. The Bloch vector subjected to amplitude damping channel transforms as $\mathbf{n}^{\rm ad}=\{ n_{1} \sqrt{1-q},n_{2} \sqrt{1-q},n_{3} (1-q) + q \}$. In order to have non zero incoherent ergotropy, the dephased state must exhibit population inversion, i.e.,  $\frac{(1-q)(1-n_{3})}{2}>\frac{(1-q)(1+n_{3})}{2}+q$, yielding two necessary constraints: (i) $n_{3}<-\frac{q}{1-q}<0$, meaning that the third Bloch component must be negative, as previously mentioned for the initial state $\rho$, (ii) $q<z<1$, where $z=\frac{\lvert n_{3} \rvert}{1+\lvert n_{3} \rvert}$, implying that the available incoherent ergotropy, if any, degrades under amplitude damping channel and vanishes in large noise strength limits ($q \to 1$). Then, it is easy to show that
\begin{align}
\mathcal W^{\rm I}(\rho^{\rm ad}) &= 
\begin{cases}
\lvert n_{3} \rvert -q(1+\lvert n_{3} \rvert) &  n_3 < 0, q<z \\[6pt]
\quad 0. & \text{otherwise}
\end{cases}
\label{incerg_amplitude_damping_single}
\end{align}
Noting Eq.~\eqref{coerg_def}, to obtain the coherent work we need to compute the energy of passive states $\pi_{\rho^{\rm ad}}$ and $\pi_{\zeta(\rho^{\rm ad})}$, where the later can vary depending on the previous constraint on non zero incoherent ergotropy. Thus, the coherent ergotropy is obtained as
\begin{align}
\mathcal W^{\rm C}(\rho^{\rm ad}) &= 
\begin{cases}
\frac{1}{2}(\lVert \mathbf{n}^{\rm ad} \rVert - \lvert n_{3} \rvert (1-q) +q) &  n_3 < 0, q<z \\[6pt]
\frac{1}{2}(\lVert \mathbf{n}^{\rm ad} \rVert + \lvert n_{3} \rvert (1-q) -q) & n_3 < 0, q \geq z \\[6pt]
\frac{1}{2}(\lVert \mathbf{n}^{\rm ad} \rVert - n_{3} (1-q) -q). & n_3 \geq 0
\end{cases}
\label{coerg_amplitude_damping_single}
\end{align}
One can analytically verify that when $n_3 < 0$ and $q<z$, then $d\mathcal W^{\rm C}(\rho^{\rm ad})/dq>0$, implying a noise-assisted growth in coherent ergotropy. However, when $q>z$ or $n_3 \geq 0$, the derivative becomes negative, and coherent ergotropy decreases monotonically. While an explicit analytic expression for the growing and freezing conditions exists, it is rather cumbersome. Instead, we explore the results numerically. Fig.~\ref{plot_amplitude_damping_single_qubit}a presents the coherent ergotropy (solid) under amplitude damping versus $q$ for several sets of Bloch vector parameters. Quantum coherence is also plotted ($\mathcal{C}/2$) obtained as
\begin{equation}
\mathcal{C}(\rho^{\rm ad}) = \sqrt{(n_1^2 + n_2^2) (1 - q)}.\label{coherence_amp}
\end{equation}
The coherent ergotropy shows a universal non-monotonic dependence on the noise strength: it increases initially, peaks around the critical point $z=\frac{\lvert n_{3} \rvert}{1+ \vert n_{3} \rvert}$ and then decreases to zero at $q=1$. We highlight the region where the $\mathcal W^{\rm C}(\rho^{\rm ad})>\mathcal W^{\rm C}(\rho)$ using shaded areas. 
The components $n_{1}$ and $n_{2}$ affect the peak symmetrically, while $n_{3}$ controls both the peak and its location versus $q$, as well as the width of the enhancement region. Fig.~\ref{plot_amplitude_damping_single_qubit}b complements these findings by a contour plot of $\Delta \mathcal W=\mathcal W^{\rm C}(\rho^{\rm ad})-\mathcal W^{\rm C}(\rho)$ versus $n_{3}$ and $q$, for $n_{1}=0.5$ and $n_{2}=0.3$. As discussed previously, a negative $n_{3}$ is required in order to have  $\Delta \mathcal W>0$. Moreover, the enhancement is confined to some critical noise strength, here $q<0.6$.

On the other hand, the coherence exhibits a strictly monotonic decay with increasing noise, approaching zero at $q=1$, as also apparent by Eq.~\eqref{coherence_amp}. As shown, $\mathcal{C}/2$ always remains above the coherent ergotropy, and the two quantities coincide either when the coherent ergotropy attains its maximal value, corresponding to the upper-bound $\mathcal{W}^{\rm C} \leq \mathcal{C}/2$, or in the strong-noise limit, where both quantities vanish.
\subsection*{B: Two-qubit system Bell diagonal States: the Role of Correlations}
We extend our findings to the case of two non interacting qubits with Hamiltonian $H=-\frac{B}{2}(\sigma_{3} \otimes \mathrm{I} + \mathrm{I} \otimes \sigma_{3})$. As the initial state of the system (in the absence of noise), we focus on the well-known class of Bell diagonal states (BDS), which by definition, are diagonal in the Bell basis~\cite{horodecki1996information,lang2010quantum}. In the Bloch representation, any BDS can be written as
\begin{align}
\rho_{\rm BDS}=\frac{1}{4}(\mathrm{I} \otimes \mathrm{I} + \sum_{i=1}^{3} c_{i} \sigma_{i} \otimes \sigma_{i}).\label{Bell-diagonal-state}
\end{align}
where $c_{i}$ are its parameters, characterizing the degree of correlation. The separability criterion of $\rho_{\rm BDS}$ is $\sum_{i} |c_{i}| \leq1$~\cite{lang2010quantum,horodecki1996information}. Throughout the rest of the manuscript, we restrict our analysis to the separable Bell-diagonal states. Obviously, the extractable work of these states depends on the relative magnitude of $c_{1}$, $c_{2}$, $c_{3}$. Throughout the rest of this work, we assume that all $c_{i}$ are non-negative, but we consider different orderings—e.g., $c_{1}>c_{2}>c_{3}$, $c_{1}>c_{3}>c_{2}$, and so on—since these permutations affect the energy spectrum of the BDS state and thus its ergotropy components. Importantly, it is easy to show that Bell diagonal states are locally passive, meaning that their local subsystems contain no extractable work under local unitaries. In fact, tracing out either qubit yields a maximally mixed state which in turn, corresponds to a thermal state at infinite temperature, hence completely passive. Therefore, any extractable work from Bell diagonal states must arise from their global properties like correlations. To investigate the connection between these correlations and ergotropy, we quantify quantum and classical correlations using geometric measures, based on the trace norm distance between the state and its closest state that lacks the specific type of correlation~\cite{modi2009unified}. We define the Geometric quantum correlation (GQC) as
\begin{align}
\mathcal Q_{\rm G}(\rho)=\lVert \rho - \rho_{\rm cl} \rVert_{1},\label{GQC1}
\end{align}
where $\lVert A \rVert_{1}={\rm Tr} \sqrt{A^{\dagger} A}$ is the trace norm, and $\rho_{\rm cl}$ denotes the closest classical state to
$\rho$. For Bell diagonal states, it is known that the set of classical states  are of the form $\rho_{\rm cl}^{i}=\frac{1}{4}(\mathrm{I} \otimes \mathrm{I} + c_{i} \sigma_{i} \otimes \sigma_{i})$~\cite{dakic2010necessary} where the closest classical state to $\rho$ is one that minimizes Eq.~\eqref{GQC1}. Thus, substituting $\rho_{\rm cl}^{i}$ into Eq.~\eqref{GQC1} yields
\begin{align}
\mathcal Q_{\rm G}(\rho)={\rm int} \{c_{i} \},\label{GQC2}
\end{align}
where {\rm int} refers to the intermediate value, i.e., the second largest element among $c_{i}$. Similarly, we define the Geometric classical correlation (GCC) as
\begin{align}
\mathcal C_{\rm G}(\rho)=\lVert \rho_{\rm cl} - \Pi_{\rho_{\rm cl}} \rVert_{1},\label{GCC1}
\end{align}
such that $\Pi_{\rho_{\rm cl}}$ shows the closest product state to $\rho_{\rm cl}$, which for Bell diagonal states is simply the maximally mixed product state $\Pi_{\rho_{\rm cl}}=\frac{1}{4} \mathrm{I} \otimes \mathrm{I}$. Therefore, the geometric classical correlation reduces to~\cite{paula2013geometric}
\begin{align}
\mathcal C_{\rm G}(\rho)={\rm max} \{c_{i}\}.\label{GCC2}
\end{align}
The following theorem demonstrates how the total work of a Bell diagonal state is linked to its quantum and classical correlations when subject to Markovian noise channels.\\
\\
\textbf{Theorem 1}: \textit{For Bell diagonal states, whose either one or both qubits undergo one of the aforementioned Markovian channels (except for amplitude damping), the total ergotropy is always equal to the average of total geometrical correlation, i.e., $\mathcal W(\rho_{\rm BDS}^{\rm q})=\mathcal W^{\rm C}(\rho_{\rm BDS}^{\rm q})+\mathcal W^{\rm I}(\rho_{\rm BDS}^{\rm q})=\mathcal I=\frac{1}{2} \left[ \mathcal C_{\rm G}(\rho_{\rm BDS}^{\rm q}) + \mathcal Q_{\rm G}(\rho_{\rm BDS}^{\rm q})\right]$}.\\

\begin{figure}[t!]
\begin{center}
\includegraphics[width=1\columnwidth]{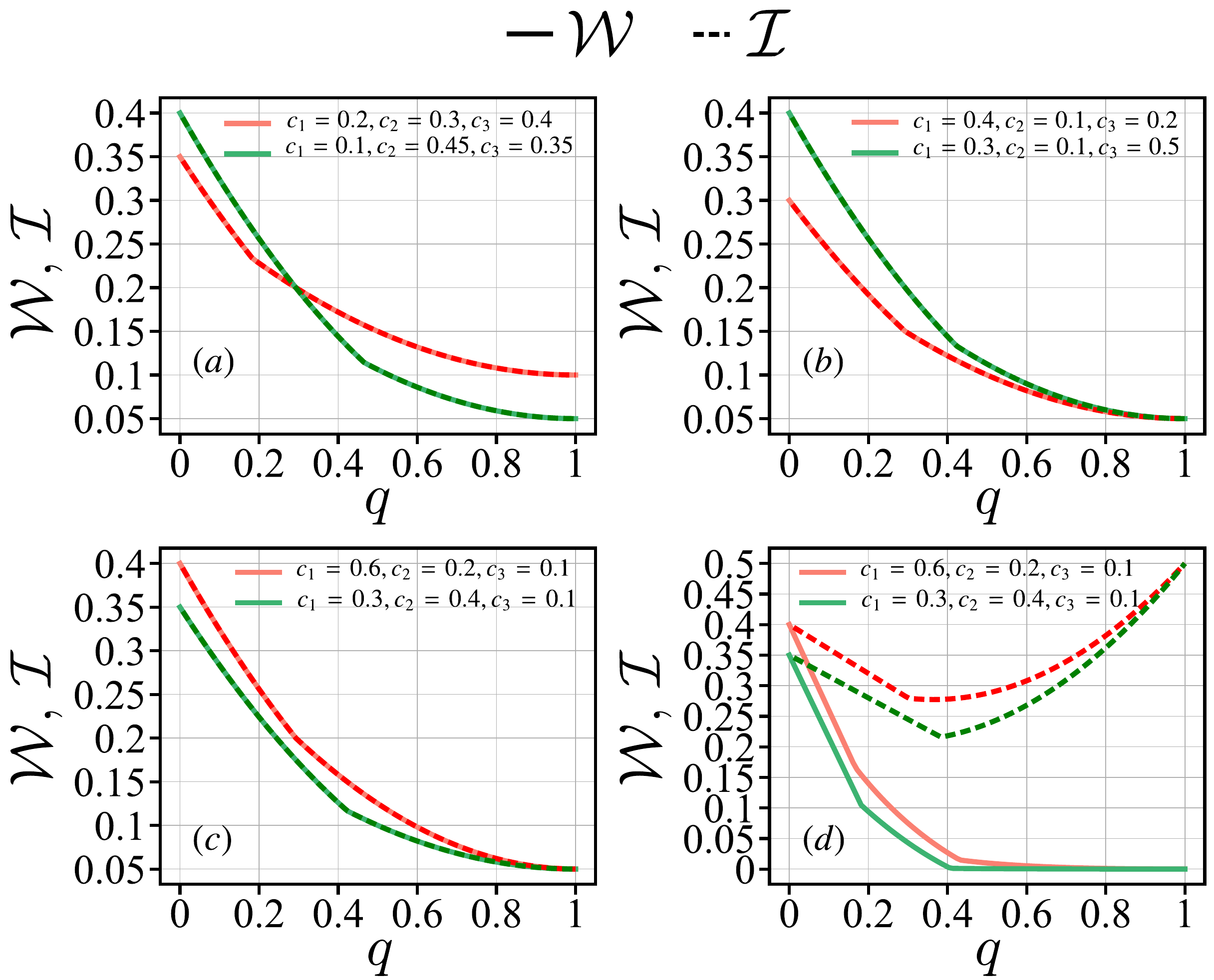}
\end{center}
\caption{Total extractable work $\mathcal W$ (solid) and the average of geometrical quantum and classical correlations $\mathcal I$ (dashed) of a two-qubit Bell-diagonal state under (a) bit flip, (b) bit-phase flip, (c) phase flip, and (d) amplitude damping. For the amplitude damping case also the average of correlations are computed using Eqs.~\eqref{GQC2} and ~\eqref{GCC2}, even though we emphasize that the formulas are not able to properly evaluate the correlation content under amplitude damping, due to non-unital nature of this channel.}
\label{four_figs_two_qubit}
\end{figure}

The proof is given in Appendix~\ref{app_B}. It is worth mentioning that the correspondence between the total extractable work and the average of quantum and classical correlations has previously been established for noiseless Bell-diagonal states~\cite{araya-sossa2019geometrical}. The above Theorem extends this relation to Bell-diagonal states subjected to Markovian noise, thereby generalizing the earlier finding. Having established the theorem, we now proceed to explore the BDS work content under different types of noise, comparing with the quantum and classical correlations, assuming that both qubits are affected by the noises.\\
\\
\textbf{Bit flip}: Applying bit flip noise to the first qubit of Eq.~\eqref{Bell-diagonal-state}, the BDS parameters changes as $C^{\rm bf}(q)=\{ c_{1}, c_{2} (1-q)^{2}, c_{3} (1-q) ^{2} \}$. The incoherent ergotropy can be obtained as
\begin{align}
\mathcal W^{\rm I}(\rho_{\rm BDS}^{\rm bf})=\frac{1}{2}c_{3} (1-q)^{2},\label{inco_erg_bit_flip}
\end{align}
which persistently decreases with the noise parameter. The coherent ergotropy, however, depends on the relative magnitude of $c_{i}$. If $c_{1} \geq c_{2} \geq c_{3} \geq 0$ then the coherent ergotropy is
\begin{align}
\mathcal W^{\rm C}(\rho_{\rm BDS}^{\rm bf})=\frac{1}{2}(c_{1} + c_{2} (1-q)^{2} - c_{3} (1-q)^{2}).\label{co_erg_bit_flip1}
\end{align}
In order to have robust (frozen) coherent ergotropy either $q=0$ or $c_{2}=c_{3}$ must hold, the former being trivial. If $c_{2} \neq c_{3}$, then $\mathcal W^{\rm C}(\rho_{\rm BDS}^{\rm bf})$ is not frozen, though it does not vanish in strong noise limits $\mathcal W^{\rm C}(\rho_{\rm BDS}^{\rm bf}) \big|_{q=1}=\frac{c_{1}}{2}$. Hence, initializing the BDS with a dominant $c_{1}$ ($c_{1} \geq c_{2} \geq c_{3} \geq 0$)—associated with coherence along the x-axis —enhances the resilience to bit flip channel. The total ergotropy is then $\mathcal W(\rho_{\rm BDS}^{\rm bf})=\mathcal W^{\rm I}(\rho_{\rm BDS}^{\rm bf})+\mathcal W^{\rm C}(\rho_{\rm BDS}^{\rm bf})=\frac{1}{2}(c_{1} + c_{2} (1-q)^{2})$. Furthermore, according to Eq.~\eqref{GQC2} and~\eqref{GCC2}, quantum and classical correlations in this case are $\mathcal Q_{\rm G}(\rho_{\rm BDS}^{\rm bf})=c_{2}(1-q)^{2}$ and $\mathcal C_{\rm G}(\rho_{\rm BDS}^{\rm bf})=c_{1}$, respectively which is consistent with theorem 1. The same results hold when $c_{2} \geq c_{1} \geq c_{3} \geq 0$, just up to swapping $\mathcal Q_{\rm G}$ and $\mathcal C_{\rm G}$. In addition, if $\min \{ c_{i} \}=c_{2}$ ($c_{1} \geq c_{3} \geq c_{2} \geq 0$ or $c_{3} \geq c_{1} \geq c_{2} \geq 0$) then $\mathcal W^{\rm C}(\rho_{\rm BDS}^{\rm bf})=\frac{1}{2} c_{1}$ which is invariant under bit flip and hence frozen. For the other two ordering relations when $\min \{ c_{i} \}=c_{1}$, eigenvalue crossing occurs such that the largest eigenvalue of $\rho_{\rm BDS}^{\rm bf}$ reshapes with $q$ (see Appendix~\ref{app_B}). Therefore, the explicit form of coherent ergotropy reshapes at some specific noise strength. For these cases, in Fig.~\ref{four_figs_two_qubit}a, we resort to numerical evaluation of the total extractable work, $\mathcal W$, which shows a perfect agreement with the average of correlation $\mathcal I$, in accordance with Theorem 1. While $\mathcal W$ decays with increasing the noise strength, it does not vanish in the strong-noise limit $q=1$ and saturates at $c_{1}/2$, which corresponds exactly to the residual coherent ergotropy in this limit, since the incoherent part being null. Furthermore, $\mathcal W$ exhibits an abrupt, yet negligible change at some specific degree of noise (depending on the state's parameters), due to eigenvalue crossing, altering the corresponding passive state and hence the coherent (and so the total) work. \\
\\
\\
\textbf{Bit-phase flip}: Under the bit-phase flip channel, the BDS parameters transform as $C^{\rm bpf}(q)=\{ c_{1}(1-q)^{2}, c_{2}, c_{3} (1-q)^{2} \}$ where $c_{2}$ remains unchanged. Given that both $c_{1}$ and $c_{2}$ contribute symmetrically to off-diagonal elements and hence coherence, the results for the bit flip channel hold true here as well, up to replacing $c_{1}$ with $c_{2}$. For this type of noise, the eigenvalues cross when $\min \{c_{i}=c_{2} \}$. In Fig.~\ref{four_figs_two_qubit}b, two instances of this case are considered for ergotropy and correlations. As in the bit flip scenario, $\mathcal W$ decreases to its ultimate value for its coherent contribution which is $c_{2}/2$, consistent with the earlier analysis.\\
\\
\textbf{Phase flip}: This channel turns the BDS parameters into $C^{\rm pf}(q)=\{ c_{1}(1-q)^{2}, c_{2}(1-q)^{2}, c_{3} \}$, leaving the third component and thus the incoherent ergotropy frozen, i.e., $\mathcal W^{\rm I}(\rho_{\rm BDS}^{\rm pf})=\frac{1}{2}c_{3}$. However, this is not the case about the coherent part. Indeed, if $\min \{c_{i} \} = c_{2}$, we have $\mathcal W^{\rm C}(\rho_{\rm BDS}^{\rm pf})=\frac{1}{2}c_{1} (1-q)^{2}$ and if $\min \{c_{i} \} = c_{1}$ then $\mathcal W^{\rm C}(\rho_{\rm BDS}^{\rm pf})=\frac{1}{2}c_{2} (1-q)^{2}$ which both decay with noise strength and disappear in $q \to 1$. For the remaining two orderings of $c_{i}$ for which $\min \{c_{i} \} = c_{3}$, the BDS state exhibit eigenvalue crossing with $q$, leading to a nontrivial dependence of the passive state and hence the coherent ergotropy on the noise strength. These cases are addressed numerically in Fig.~\ref{four_figs_two_qubit}c, which resemble those obtained for the bit-flip and bit-phase-flip channels. The key difference is that, under phase flip, the residual extractable work in the high-noise limit ($q=1$) originates solely from the frozen incoherent ergotropy and not coherent part, since the later is not robust against phase flip the coherent, as discussed above.\\
\\
\textbf{Depolarizing channel}: All three components of the state are uniformly attenuated under this channel, resulting in $C^{\rm dc}(q)=\{ c_{1}(1-q)^{2}, c_{2}(1-q)^{2}, c_{3}(1-q)^{2} \}$. As $q$ approaches 1, the state converges to the maximally mixed state, which is energetically inactive under unitary operations. Therefore, neither coherent nor incoherent ergotropy exhibits freezing behavior under this channel.\\
\\
\textbf{Amplitude damping}: This channel stands apart from all previously considered ones due to its non-unital character; it does not preserve the identity operator. When applied to a Bell-diagonal state, the off-diagonal parameters transform as $c_{1,2}\to c_{1,2} (1-q)$, whereas the third component, $c_{3}$ evolves asymmetrically. In particular, although the ground and excited states of the unperturbed BDS Eq.~\eqref{Bell-diagonal-state} have the same population, they shift to $\frac{1}{4}(1+c_{3}(1-q)^{2} + (1+q)^2)$ and $\frac{1}{4}(1+c_{3}(1-q)^2)$, respectively. This imbalance destroys the symmetry required to preserve the Bell-diagonal form, thereby invalidating Theorem 1. Moreover, it is not difficult to show that under amplitude damping, the incoherent ergotropy changes as
\begin{figure}[h!]
\begin{center}
\includegraphics[width=0.6\columnwidth]{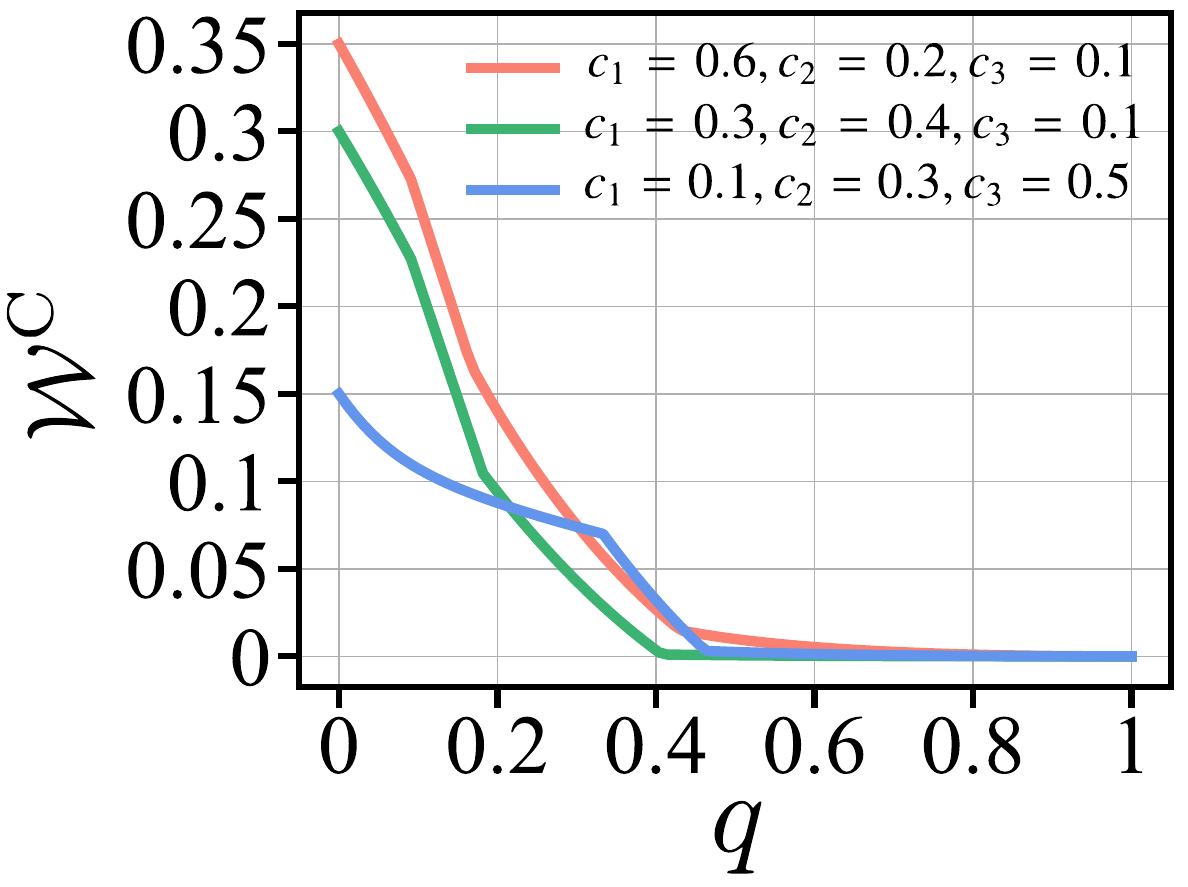}
\end{center}
\caption{Coherent contribution of extractable energy under amplitude damping.}
\label{coherent_erg_amplitude_damping_bell_daigonal}
\end{figure}

\begin{align}
\mathcal W^{\rm I}(\rho_{\rm BDS}^{\rm ad}) &= 
\begin{cases}
\frac{1}{2}(-1+q)(c_{3}(-1+q)+q) & q<\frac{c_{3}}{1+c_{3}} \\[6pt]
\quad 0 & \text{otherwise}
\end{cases}
\label{incerg_amplitude_damping}
\end{align}

\begin{figure*}[t!]
\begin{center}
\includegraphics[width=\textwidth]{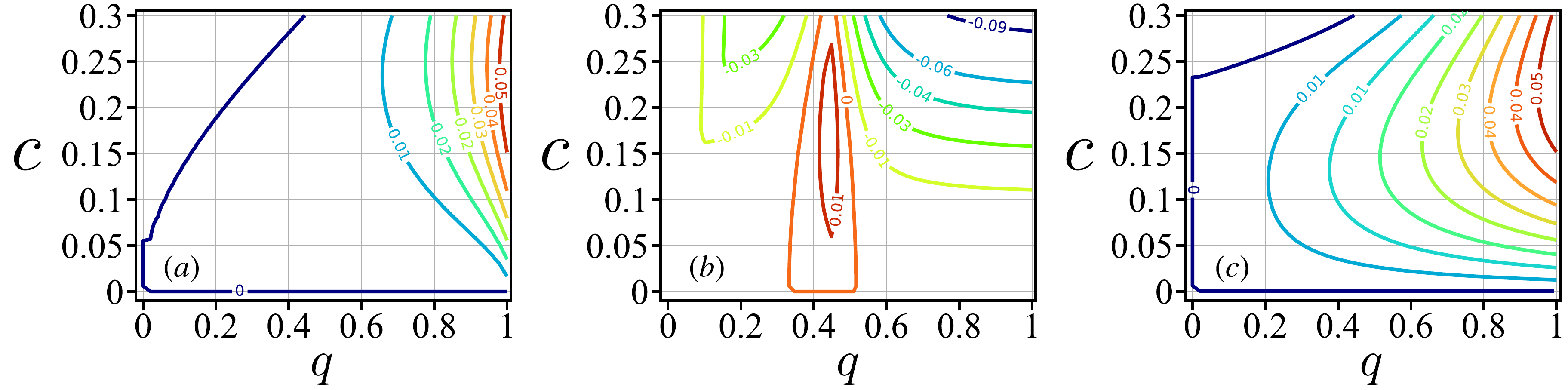}
\end{center}
\caption{Difference in coherent ergotropy with and without noise given by $\Delta \mathcal{W}^{\rm C}=\mathcal{W}^{\rm C}(q)-\mathcal{W}^{\rm C}(0)$ for the classical-quantum state Eq.~\eqref{classical-quantum state} under (a) bit flip, (b) amplitude damping and (c) phase flip, versus coherence parameter and noise strength. Similar results to the bit flip can be obtained for bit-phase flip by selecting complex number for off-diagonal elements of the state. The other parameters are $p=0.5$, $a=0.1$.}
\label{contour_classical_quantum}
\end{figure*}
which is not frozen. As for coherent ergotropy, given that several eigenvalue crossings can happen in various orderings of $c_{i}$, we only resort to numerical analysis, where no increasing behavior is observed for coherent ergotropy under amplitude damping, as shown in Fig~\ref{coherent_erg_amplitude_damping_bell_daigonal}. 

In Fig.~\ref{four_figs_two_qubit}d, the total ergotropy under amplitude damping is seen to decay monotonically, reaching zero at $q=1$ because neither the coherent nor the incoherent component is preserved under this channel. We also include the correlations average calculated using Eqs.~\eqref{GQC2} and ~\eqref{GCC2}, however, we note that since the Bell diagonal structure of the state is not preserved after amplitude damping, the considered formulas fail to properly capture the correlation content of this state. As illustrated by Fig.~\ref{four_figs_two_qubit}d it is evident that the average of these formulas does not reflect the total extractable work, rendering the Theorem 1 inapplicable under amplitude damping.

\subsection*{C: Two-qubit separable state with local coherence}
In the preceding subsection, we examined Bell-diagonal separable states $(|c_1| + |c_2| + |c_3| \leq 1)$, whose maximally mixed marginals preclude local coherence. We now extend the analysis to separable states with local coherence. We begin by a simple classical--quantum state
\begin{equation}
\rho_{\rm cq} = p\, \ket{\rm g}\bra{\rm g} \otimes \rho(a,c) + (1-p)\, \ket{\rm e}\bra{\rm e} \otimes \rho(a,0),\label{classical-quantum state}
\end{equation}
in which $0\leq p \leq 1$ and
\begin{equation}
\rho(x,y) =
\begin{pmatrix}
x & y \\
y^* & 1-x
\end{pmatrix},\label{rho(x,y)}
\end{equation}
where for simplicity, we have assumed that both local states of the second subsystem share identical populations, but only one exhibits coherence. Though this state remains separable, it can possess nonzero \textit{right quantum discord}, when performing measurements on the second subsystem, provided $[\rho(a,c),\, \rho(a,0)] \neq 0$~\cite{piani2008no,dakic2010necessary}. Such a genuine quantum correlation beyond entanglement has been identified to be essential for energy-storage processes in quantum batteries~\cite{cruz2022quantum}.

A finite coherent contribution to the ergotropy arises only when the state contains nonzero coherence, i.e., $c\neq 0$. Indeed, recalling the noninteracting Hamiltonian $H$, the dephased state $\zeta(\rho)$ is obtained by simply eliminating the off-diagonal terms of $\rho$, without altering its populations. Hence, for $c=0$, the dephased and original states coincide, implying that their passive states are also identical, yielding vanishing coherent ergotropy according to Eq.~\eqref{coerg_def}. Furthermore, since the locally applied Kraus operators considered here cannot generate coherence~\cite{hu2016channels}, the coherent ergotropy remains zero under each noise channel. In contrast, when $c\neq 0$, the initial local coherence gives rise to a finite coherent ergotropy, which, under certain noise channels, can even increase. 

Since deriving analytical conditions for the enhancement of the coherent part of ergotropy under noise quickly becomes cumbersome, we resort to numerical analysis for the following discussion. For the class of states defined above, we find that the coherent ergotropy can increase under the action of the considered channels within specific parameter regimes. 

Fig.~\ref{contour_classical_quantum} displays contour plots of the change in coherent ergotropy, defined as $\Delta \mathcal{W}^{\rm C}=\mathcal{W}^{\rm C}(q)-\mathcal{W}^{\rm C}(0)$ versus coherence parameter $c$ and noise strength $q$ for bit flip, amplitude damping and phase flip. Similar results to those obtained for the bit flip channel can also be achieved for the bit-phase-flip by choosing a complex value of $c$. Here, $\mathcal{W}^{\rm C}(0)$ and $\mathcal{W}^{\rm C}(q)$ refer to the coherent work of the initial classical–quantum state Eq.~\eqref{classical-quantum state} and of the corresponding state after the action of noise, respectively. Under all noises, regions appear where the coherent work remains frozen ($\Delta \mathcal{W}^{\rm C}=0$) or even increases ($\Delta \mathcal{W}^{\rm C}>0$). The dependence on the parameters, however, differs markedly between the channels. Under bit-flip (Fig.~\ref{contour_classical_quantum}a), enhancement of $\mathcal{W}^{\rm C}$ occurs only for sufficiently large noise strengths, with $\Delta \mathcal{W}^{\rm C}$ growing as $q$ increases. The variation with respect to coherence parameter is nearly non-monotonic, in particular with smaller $q$, where $\Delta \mathcal{W}^{\rm C}$ rises from zero at $c=0$, reaches a maximum at intermediate coherence value, and slightly decreases for larger $c$. In contrast, for the amplitude-damping channel, nearly all nonzero values of coherence parameter can yield a positive $\Delta \mathcal{W}^{\rm C}$, provided the noise strength lies within a moderate range ($q \approx 0.3-0.5$). This enhancement region becomes narrower as $c$ increases (Fig.~\ref{contour_classical_quantum}b).

In the present analysis, we have considered the local Hamiltonian aligned with the computational ($ \sigma_z $) basis. Within this choice, the phase-flip channel degrades the off-diagonal elements of the density matrix in the same basis, leading to a monotonic reduction of coherence and, consequently, of the coherent ergotropy. However, as previously shown for the single-qubit case, the behavior of coherent ergotropy under phase flip noise can change considerably when the energy basis is modified. In particular, when the Hamiltonian is defined $H=\frac{1}{2} (\sigma_{x}^{1}+\sigma_{x}^{2})$, the phase-flip channel no longer acts as pure dephasing with respect to the energy eigenbasis but rather introduces population--coherence mixing. This mixing can convert parts of the initial population imbalance into off-diagonal terms, resulting in a nontrivial enhancement of the coherent ergotropy even under dephasing noise, as shown in Fig.~\ref{contour_classical_quantum}c. Similar to the case of bit flip, under phase flip channel also $\Delta \mathcal{W}^{\rm C}$ exhibits non-monotonic dependence with respect to coherence parameter. However, phase flip results in a considerably broader range of $q$ where $\Delta \mathcal{W}^{\rm C} > 0$. This enhancement is most pronounced for higher values of $c$ and stronger noise strengths.

\begin{figure*}[t]
\begin{center}
\includegraphics[width=\textwidth]{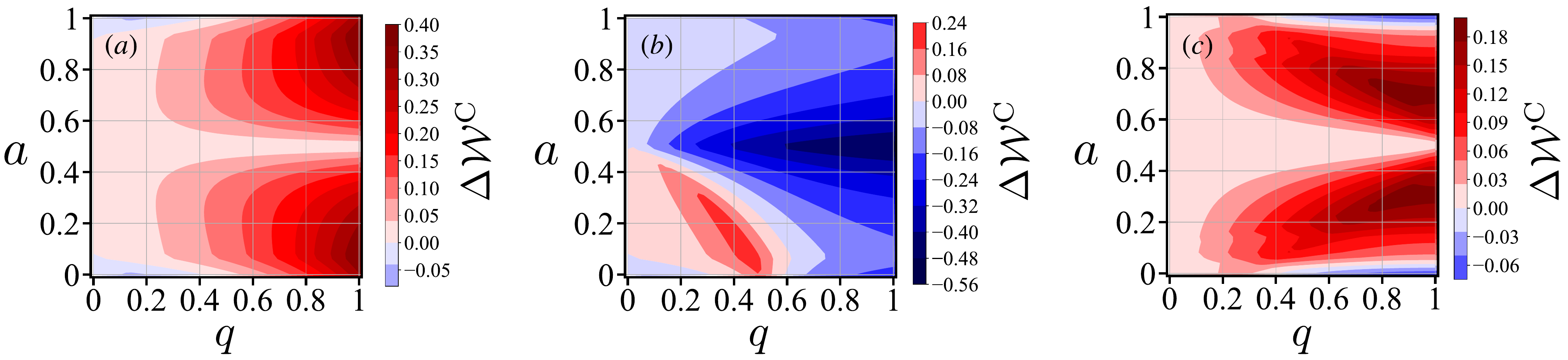}
\end{center}
\caption{Similar to Fig.~\ref{contour_classical_quantum} for the separable state Eq.~\eqref{quantum-quantum} versus population parameter and noise strength. The other parameters are $p=0.5$, $c=0.3$, and $d=0.2$.}
\label{contour_quantum_quantum}
\end{figure*}

We now extend our analysis to a more general class of separable states in which both subsystems possess local coherence. The state is defined as
\begin{equation}
    \rho_{S} = p\, \rho(a,c) \otimes \rho(a,d) + (1-p)\, \rho(a,d) \otimes \rho(a,c),\label{quantum-quantum}
\end{equation}
where each local density matrix is given by Eq.~\eqref{rho(x,y)}. The local populations are chosen to be identical for both subsystems, while the coherence amplitudes $c$ and $d$ may differ. Fig.~\ref{contour_quantum_quantum} illustrates $\Delta \mathcal{W}^{\rm C}$ as a function of the noise strength $q$ and the population parameter $a$, for fixed values of $c$ and $d$. Under bit-flip noise, nearly the entire parameter range leads to an enhancement of coherent work, with the increase becoming more pronounced at higher noise strengths---reaching values up to $\Delta \mathcal{W}^{\rm C} \simeq 0.36$, as shown by Fig.~\ref{contour_quantum_quantum}a. The dependence on the population parameter exhibits a distinct horizontal band around $a \approx 0.5$, corresponding to locally balanced populations, where the coherent work remains frozen ($\Delta \mathcal{W}^{\rm C} \approx 0$). This stems from degenerate dephased states, whose the energy of the corresponding passive states vanishes (see Appendix~\ref{app_C}). The results are symmetric with respect to this band, and significant enhancement occurs on both sides except for very small or very large $a$, where $\Delta \mathcal{W}^{\rm C}$ slightly decreases.
However, for the amplitude-damping channel (Fig.~\ref{contour_quantum_quantum}b), the enhancement of coherent work is confined to a narrower region of the parameter, occurring only for moderate noise strengths ($q \lesssim 0.6$) and smaller ground-state populations ($a < 0.5$). Within this region, the coherent work can still increase substantially, reaching values up to $\Delta \mathcal{W}^{\rm C} \simeq 0.22$ for appropriately tuned parameters. analogous results can be obtained for the bit-phase flip channel by choosing complex off-diagonal elements in Eq.~\eqref{quantum-quantum}. To maintain brevity, the corresponding results are not included here.

Similar results to the bit-flip channel are obtained for the phase-flip noise, featuring two symmetric regions of coherent ergotropy enhancement around $a=0.5$, with the enhancement becoming stronger as the degree of noise is increased. Using the same argument is that given in Appendix~\ref{app_C}, the emergence of the frozen band can be justified also for this noise. Yet, the domain of which $\Delta \mathcal{W}^{\rm C}>0$ in the $a-q$ plane is comparatively more limited than bit flip, most noticeably at low noise strengths, as illustrated in Fig.~\ref{contour_quantum_quantum}c.

In Appendix~\ref{app_D} we demonstrate that depolarizing channels can also lead to the enhancement of coherent ergotropy extracted from the state in Eq.~\eqref{quantum-quantum}, when the system evolves under an interacting Hamiltonian.

So far, we have considered the separable state Eq.~\eqref{quantum-quantum} constructed as a permutation of two local states with identical populations and different coherences. To demonstrate that this choice is not restrictive, in Appendix~\ref{app_E} we extend the analysis to general separable two-qubit states with independently randomized local parameters.

\section{Generalization to multipartite states}
\label{multi}
In this section, we proceed by extending our analysis to multipartite systems to investigate how the enhancement of coherent ergotropy under noise scales with the size of system. We generalize the separable state introduced previously in Eq.~\eqref{quantum-quantum} to an $N$-body state as
\begin{equation}
\rho_S^{(N)} = \frac{1}{N!}\,\mathcal{S}\!\left[\bigotimes_{i=1}^{N} \rho(a, c_i)\right],\label{multipartite_state}
\end{equation}
where $\mathcal{S}[\cdot]$ denotes the symmetrization operation over all possible permutations of the local states. For instance, in the case of three qubits, the state is an equally weighted mixture of all $6$ possible permutations of the local states $\rho(a, c_1)$, $\rho(a, c_2)$, and $\rho(a, c_3)$ such as $\rho(a, c_1) \otimes \rho(a, c_2) \otimes \rho(a, c_3)$, $\rho(a, c_1) \otimes \rho(a, c_3) \otimes \rho(a, c_2)$ and so on, accounting for all distinct orderings of the local coherent states in the tensor product. Once again, we assume that each subsystem shares the same ground-state population parameter $a$, while the local coherence amplitudes vary as $c_i = c_0 + i\,\delta$, allowing the subsystems to be distinguished by their local degree of coherence.
To characterize how the enhancement of coherent ergotropy scales with the system size, we consider two quantities. 
The first is the \textit{maximum enhancement}, defined as
\begin{equation}
\Delta \mathcal{W}^{\rm C}_{\max} = \max_q \, \mathcal{W}^{\rm C}(q),\label{DeltaW_max}
\end{equation}
which captures the largest increase of coherent ergotropy over the range of noise strengths $q$.
The second quantity is the \textit{area of enhancement},
\begin{equation}
\mathcal{A}_{\rm p} = \int_{\Delta \mathcal{W}^{\rm C}(q) > 0} \! \Delta \mathcal{W}^{\rm C}(q)\, dq,\label{enhancement_area}
\end{equation}
which measures the total region of noise strengths for which the coherent ergotropy exceeds its noiseless value.

\begin{figure}[h]
\begin{center}
\includegraphics[width=\columnwidth]{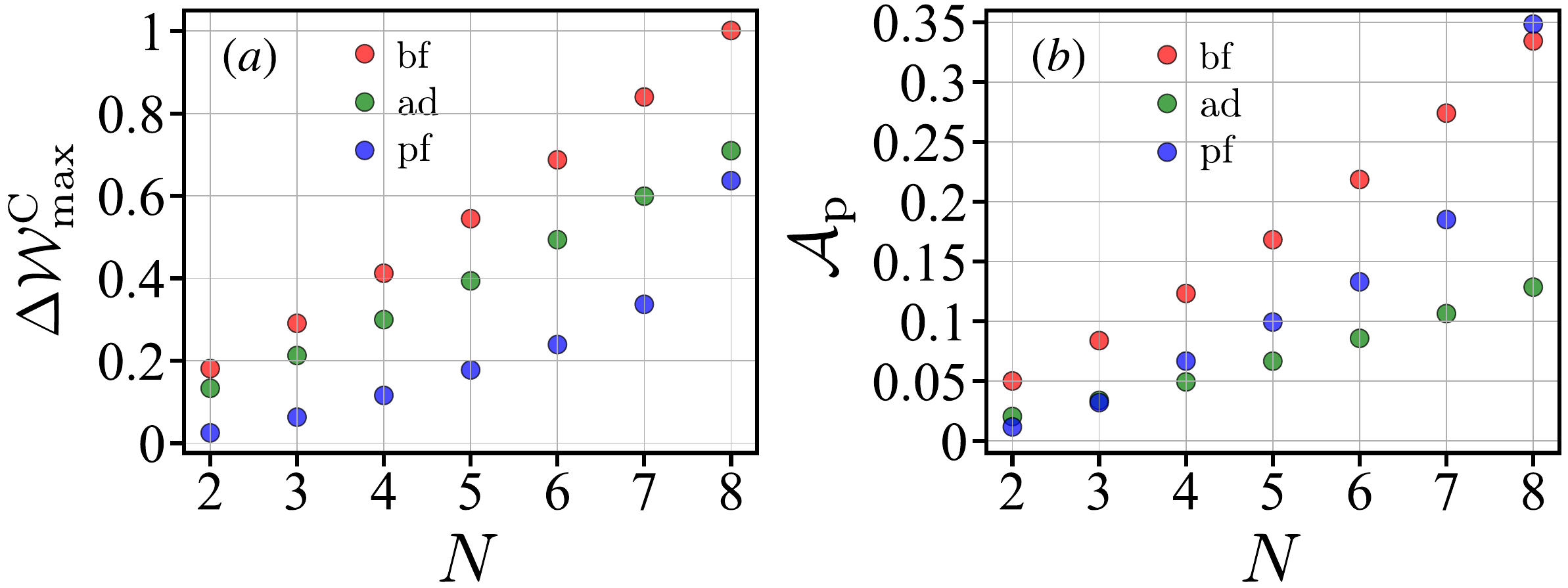}
\end{center}
\caption{Scaling of maximum enhancement of coherent ergotropy and the area of enhancement given by Eq.~\eqref{DeltaW_max} and~\eqref{enhancement_area}, respectively, with the number of qubits,. Here, $a=0.2$ and $c_i = c_0 + i\,\delta$ where $c_0=0.1$ and $\delta=0.02$.}
\label{scaling_with_N}
\end{figure}

\begin{figure}[h]
\begin{center}
\includegraphics[width=\columnwidth]{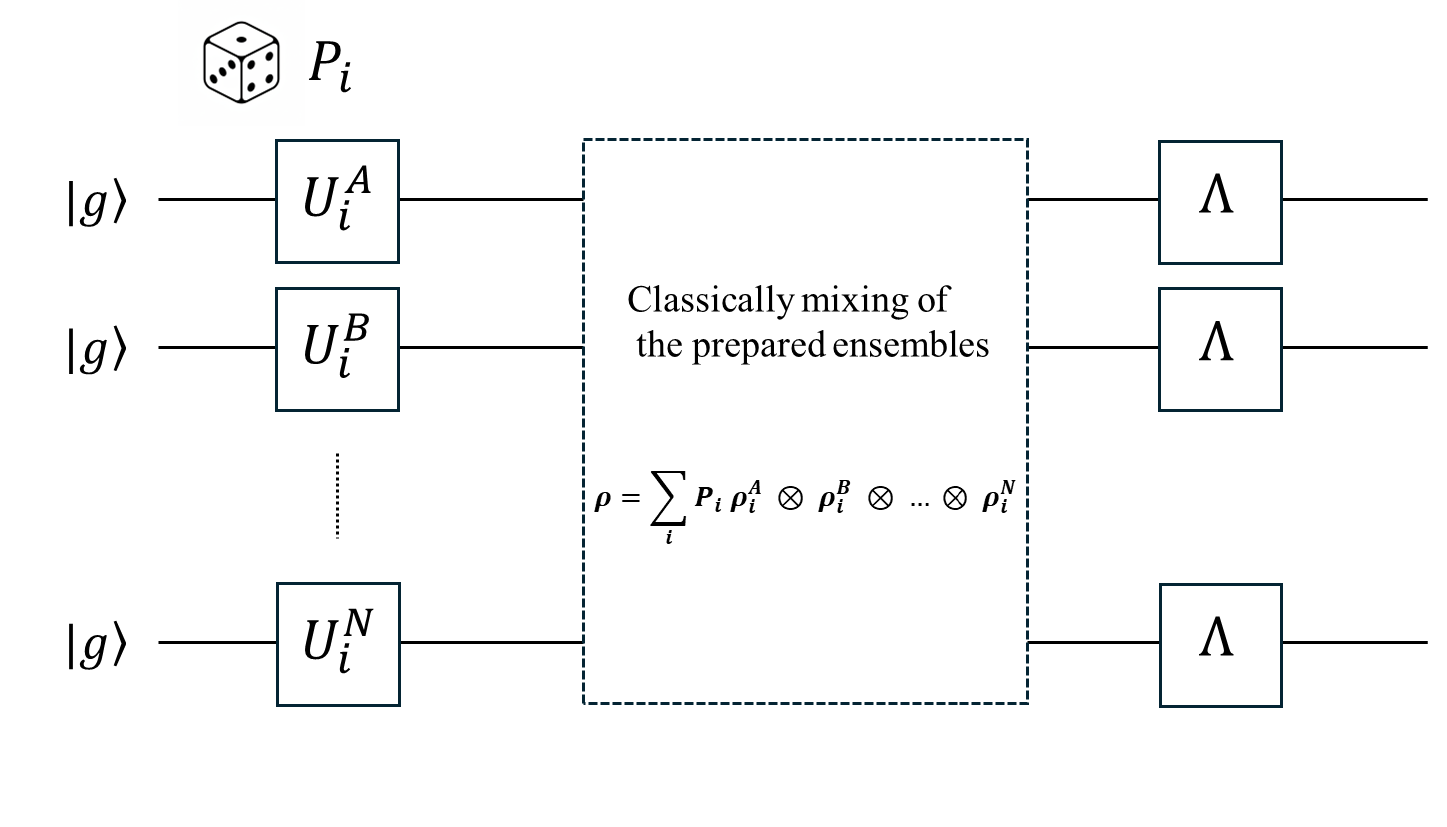}
\end{center}
\caption{Schematic representation of the protocol, from preparation (charging) to noisy evolution. Each qubit is initialized in the ground state in the ground state ($\ket{g}$) of its local Hamiltonian. Local randomized unitaries with associated probabilities $P_{i}$ are then applied, followed by a classical mixing process that yields a general separable state possessing local coherence. Finally, the prepared state is subjected to local noise channel.}
\label{sketch}
\end{figure}

Fig.~\ref{scaling_with_N} illustrates how these quantities scale with the number of qubit $N$, adopting Hamiltonians that are constructed as direct generalizations of those previously defined in the two-qubit analysis for each noise channel. It should be emphasized that no parameter optimization is performed in this analysis and $\Delta \mathcal{W}^{\rm C}_{\max}$ is obtained by maximizing solely with respect to the noise strength $q$, while all other parameters remain fixed. As shown, both $\Delta \mathcal{W}^{\rm C}_{\max}$ and $\mathcal{A}_{\rm p}$ display a clear tendency to increase with the number of qubits for all noise types, underscoring the cooperative role of multipartite accumulated coherence in sustaining noise-induced coherent ergotropy. The most significant enhancement occurs under bit-flip noise (red markers), where both quantities exhibit an almost linear scaling with $N$. For the amplitude-damping channel (green markers), both measures also increase with $N$, but with a gentler slope, consistent with Fig. 6(b), where the magnitude and region of enhancement are more limited than bit flip noise. Nevertheless, $\Delta \mathcal{W}^{\mathrm{C}}_{\max}$ for amplitude damping consistently remains above that of the phase-flip channel across all sizes (Fig.~\ref{scaling_with_N}a), reflecting a more efficient utilization of quantum coherence and population difference to increase coherent contribution of extractable energy.
Regarding the area of enhancement, however, the phase flip channel shows a distinct behavior, as observed by Fig.~\ref{scaling_with_N}b. While $\mathcal{A}_{\rm p}$ for phase flip starts below the other two channels for small systems, its slope increases substantially with $N$, surpassing amplitude damping at $N=4$ and even bit flip at $N=8$, demonstrating a nontrivial multipartite enhancement that can counteract the destructive effects of phase-flip noise on coherence.
Finally, to provide a simple concrete sketch of realization, we present a schematic of the overall protocol from charging or preparation to exposure to noise in Fig.~\ref{sketch} which can be implemented in various platforms such as super conducting qubits~\cite{kjaergaard2020superconducting}, nitrogen-vacancy (NV) centers in diamond~\cite{doherty2013nitrogen} or trapped ions~\cite{moses2023race,haffner2008quantum}.

The protocol begins with each qubit initialized in the ground state of its local Hamiltonian, for instance $H_{0}=\ket{\rm e}\bra{\rm e}$, representing an initially empty battery. Subsequently, local unitary operations $U_{i}^{A}, U_{i}^{B},...$ are applied to each qubit with associated probabilities $P_{i}$. For instance, these unitaries can be of the form $U_{i}=\exp(-i \sigma_{\rm x} \theta_{i})$, where $\theta_{i}$ controls the degree of local excitation or coherence introduced.
After the local operations, a classical mixing (or “forgetting”) step is applied, where the specific preparation in each round is not tracked. This process generates a probabilistic ensemble of tensor-product states, resulting in a separable state of the form $\rho=\sum_{i} P_{i} \rho_{i}^{A} \otimes \rho_{i}^{B} \otimes ...\otimes \rho_{i}^{N}$, where the local subsystems can retain coherence in the energy eigenbasis. Finally, local noise channels $\Lambda$ are applied to each qubit, modeling the interactions with the environment as discussed in the main text. This protocol thus provides a clear and experimentally feasible framework to study the interplay between initial coherence and noise-assisted work extraction in quantum batteries.

One may raise a concern regarding the choice of separable initial states, which are not generated through collective charging operations typically associated with quantum speed-up in many-body quantum batteries. It is important to emphasize that our focus here lies primarily on the resilience of stored energy under noise, rather than on charging-time optimization. Moreover, the considered separable states, while lacking entanglement, can still possess nonclassical correlations such as quantum discord, which appear to be essential for optimal energy extraction in quantum batteries~\cite{giorgi2015correlation,cruz2022quantum}. On the other hand, it remains possible to explore entangled states endowed with local coherence, which could in principle combine both faster charging advantages and noise resilience. In Appendix~\ref{app_G}, we provide a concrete example of such a scenario and show numerically that noise-assisted enhancement of coherent ergotropy can indeed occur in the presence of entanglement. This demonstrates that entanglement is not a hindrance to the enhancement of coherent ergotropy, and that both features can coexist within the same protocol. Consequently, one can envision quantum battery schemes that simultaneously benefit from charging speed-up due to entangling operations and enhanced robustness of the stored energy under noise. A systematic investigation of such correlated charging and storage mechanisms is left for future work.
\section{Conclusion}
\label{conc}
In this paper, we investigated the influence of Markovian noise on extractable work in quantum systems, with a particular focus on the partitioning of ergotropy into coherent and incoherent contributions in single- and two-qubit systems. For the single-qubit scenario, we found that incoherent work generally diminishes under noise, except in the case of phase flip, where the incoherent ergotropy remained invariant. This is due to the fact that the incoherent contribution is primarily governed by the third component of the Bloch vector, $n_{3}$, which remains unaffected by phase flip. Conversely, under specific conditions, coherent ergotropy can either be preserved or enhanced. Notably, in the presence of bit flip or bit-phase flip noise, we demonstrated that stronger noise can enhance the coherent extractable work by suitably tuning the components of the Bloch vector. However, this preservation is not observed under phase flip or depolarizing channels, as these either degrade coherence or drive the system to a maximally mixed state. Additionally, amplitude damping exhibited more intricate behavior, with certain noise strength regimes leading to a non-monotonic increase in coherent ergotropy. In addition, we established an upper bound showing that coherent ergotropy is limited by one half of the system’s quantum coherence.

Moving to the case of two-qubit systems, where both qubits are affected by noise, we analyzed two distinct families of separable states. In the first case, corresponding to Bell-diagonal states whose marginals are completely passive and hence devoid of local coherence, we proved that the total extractable work under noise is fundamentally linked to the geometric quantum and classical correlations. This result is particularly important, as the Bell-diagonal state is locally passive, implying that any extractable work must come from the non-local features of the state. For this state, while no increasing behavior for available work was observed, we showed that either incoherent or coherent ergotropy could be frozen or have non vanishing residual value in the limit of strong noise, depending on the noise type and the ordering of the Bell diagonal state parameter. For instance, we showed that for bit-flip noise, the extractable coherent work was maximized when the first component of the Bell state (associated with the x-axis coherence) was given more weight. In this case, coherent work can be frozen or remain non-zero in the high-noise limit, specifically at $q=1$, where the work approaches $c_{1}/2$, while for bit-phase flip this role is given to the second component $c_{2}$.

In the second case, representing separable states with local coherence, we demonstrated that coherent ergotropy can be enhanced under all standard noise channels, including depolarizing and phase flip noise that typically suppress coherence, provided that the system Hamiltonian and interaction structure are suitably selected. The enhancement is particularly pronounced and extended over a broader interval of noise strengths for flip-type channels. 

Although the considered states are separable, they still possess genuine quantum correlations in the form of quantum discord, which has been shown to serve as a resource for quantum batteries~\cite{cruz2022quantum,bai2025quantum}.

Extending the analysis to multipartite systems, we showed that noise-assisted enhancement not only persists but exhibits a collective amplification as the system size increases, as evidenced by the scaling of both the maximum enhancement and area of enhancement over noise strength.

This work challenges the traditional view that noise is inherently detrimental to quantum systems, particularly in the context of quantum batteries. Instead, By strategically adjusting the system parameters and noise strength, noise can be harnessed to either enhance the stored energy or leave the available energy intact, by modulating the coherence and correlations. These findings could open up new possibilities for developments of quantum batteries, where noise becomes a valuable resource rather than a hindrance. 

Although the present work is primarily devoted to separable initial states, such states can still host genuine quantum correlations in the form of quantum discord, which has been identified as a relevant resource for quantum batteries~\cite{von2020two,zou2024spatially,gaikwad2022simulating}. Nevertheless, we also examined a representative example of an entangled state, for which entanglement and noise-assisted enhancement of coherent ergotropy coexist. This shows that entanglement does not hinder the enhancement mechanism discussed here. Consequently, the advantages of fast charging enabled by entangling operations and the noise-assisted enhancement of stored energy can, in principle, be combined within the same quantum battery protocol.

From an operational perspective, our results can be naturally mapped onto leading experimental platforms proposed for realization of quantum batteries, including superconducting circuits~\cite{hu2022optimal,dou2023superconducting}, trapped-ion systems~\cite{zhang2025single,wen2025dicke}, and spin arrays~\cite{le2018spin,dou2022cavity,yao2022optimal}. In these platforms, energy storage and extraction rely on coherent control using microwave, laser, or magnetic-field driving, while unavoidable environmental interactions or control imperfections give rise to photon loss, spontaneous emission, flip errors, and dephasing.  These physical decoherence mechanisms are well captured by the Markovian noise channels considered in this work, providing a realistic description of noise acting during the charging or storage stages of quantum batteries.

Within these platforms, the preparation of local quantum coherence—identified in this work as a key resource for enhancing coherent ergotropy—is achieved through standard high-fidelity coherent control protocols routinely used for state initialization and manipulation~\cite{yoshioka2023active,somoroff2023millisecond,siddiqi2021engineering,myatt2000decoherence,haffner2008quantum,huang2024high}. Furthermore, the Hamiltonian and interaction engineering required to activate noise-assisted enhancement, particularly under depolarizing noise, is experimentally accessible through controllable two-body couplings and effective interaction modulation. These established control techniques provide a concrete physical route for implementing the mechanisms identified here, translating our theoretical results into experimentally relevant design principles for quantum batteries operating in noisy environments.

\appendix

\section{Proof of the upper bound Eq.~\eqref{upper_bound_coherence_bf}}
\label{app_A}
In this appendix, we prove the upper bound relation
\begin{equation}
W_c \le \tfrac{1}{2}\mathcal{C},
\end{equation}
as presented in Eq.~\eqref{upper_bound_coherence_bf} of the main text. We prove this relation for a generic single-qubit state under arbitrary noise, showing that it holds independently of the specific channel.

The coherent part of the ergotropy under each markovian channel can be written as
\begin{equation}
\begin{split}
\mathcal{W}^{\rm C} =& \frac{1}{2}(\lVert \mathbf{n}(q) \rVert - \lvert n_{3}(q) \rvert ) =\\
&\frac{1}{2}\!\left[\sqrt{\,n_1^2(q) + n_2^2(q) + n_3^2(q)\,} - |n_3(q)|\right],
\label{Wc_bf_app}
\end{split}
\end{equation}
while the quantum coherence of the state, quantified by the $l_1$-norm measure, reads
\begin{equation}
\mathcal{C} = \sqrt{\,n_1^2(q) + n_2^2(q)\,},
\label{C_app}
\end{equation}

To show that $W_c \le \tfrac{1}{2}\mathcal{C}$, we recall the elementary inequality valid for all real $\Theta \geq 0$ and $\Xi \geq 0$:
\begin{equation}
\sqrt{\Theta + \Xi^2} - \Xi \leq \sqrt{\Theta}.\label{inequality}
\end{equation}

By identifying
\[
\Theta = n_1^2(q) + n_2^2(q), \qquad \Xi = |n_3(q)|,
\]
we can rewrite Eq.~\eqref{Wc_bf_app} as
\[
2W_c = \sqrt{\Theta + \Xi^2} - \Xi.
\]
Using the inequality~\eqref{inequality}, we then obtain
\[
2W_c \le \sqrt{\Theta} = \sqrt{n_1^2(q) + n_2^2(q)} = \mathcal{C},
\]
which immediately gives
\begin{equation}
\mathcal{W}^{\rm C} \leq \frac{1}{2}\mathcal{C},\label{proof_done}
\end{equation}
which completes the proof. $\blacksquare$

The equality holds when $|n_3(q)|=0$. For example, under bit flip channel, $n_{3}(q)=n_{3} (1-q)$. Then $\mathcal{W}^{\rm C} = \frac{1}{2}\mathcal{C}$ if $n_{3}=0$ or $q=1$.

\section{Proof of theorem 1}
\label{app_B}
In order to prove the theorem 1, we consider the Bell-diagonal state given by Eq.~\eqref{Bell-diagonal-state}, where one or both subsystems are subjected to one of the Markovian noises (except for the amplitude damping)
\begin{align}
\rho_{\rm BDS}(q)=\frac{1}{4}(\mathrm{I} \otimes \mathrm{I} + \sum_{i=1}^{3} c_{i}(q) \sigma_{i} \otimes \sigma_{i}).\label{BDS_under_noise}
\end{align}
where depending on each noise, the BDS parameters can evolve as $c_{i}(q)=c_{i} (1-q)$ ($i=1, 2, 3$) if only one subsystem is affected by noise or $c_{i}(q)=c_{i} (1-q)^{2}$ if both qubits are affected, or can remain unchanged $c_{i}(q)=c_{i}$. It is easy to show that the internal energy of $\rho_{\rm BDS}(q)$ is zero ${\rm Tr}\left[ \rho_{\rm BDS}(q) H \right]=0$ where $H=-\frac{B}{2}(\sigma_{3} \otimes \mathrm{I} + \mathrm{I} \otimes \sigma_{3})$ (in which we set $B=1$). So the total ergotropy of $\rho_{\rm BDS}(q)$ depends only on its passive state energy. The eigen values of $\rho_{\rm BDS}(q)$ are given by
\begin{align}
\lambda_1 &= \frac{1}{4} \left( 1 - c_1(q) - c_2(q) - c_3(q) \right),\label{eig1_bell_diagonal} \\
\lambda_2 &= \frac{1}{4} \left( 1 + c_1(q) + c_2(q) - c_3(q) \right),\label{eig2_bell_diagonal} \\
\lambda_3 &= \frac{1}{4} \left( 1 + c_1(q) - c_2(q) + c_3(q) \right),\label{eig3_bell_diagonal} \\
\lambda_4 &= \frac{1}{4} \left( 1 - c_1(q) + c_2(q) + c_3(q) \right). \label{eig4_bell_diagonal}
\end{align}
Noting that $H$ has two non zero eigen energies as $\pm 1$, we only need to specify the largest and smallest eigenvalues to obtain the passive state energy which can be simply given by
\begin{align}
\lambda_{\rm max} &= \frac{1}{4} \left(1 + {\rm max} \{ c_{i}(q) \} + {\rm int} \{ c_{i}(q) \} - {\rm min} \{ c_{i}(q) \} \right), \\
\lambda_{\rm min} &= \frac{1}{4} \left(1 - {\rm max} \{ c_{i}(q) \} - {\rm int} \{ c_{i}(q) \} - {\rm min} \{ c_{i}(q) \} \right),
\end{align}
regardless of the ordering relations of $c_{i}$. For example, if $c_{1}(q) \geq c_{2}(q) \geq c_{3}(q)$, then $\lambda_{\rm max}=\lambda_2$ and $\lambda_{\rm min}=\lambda_1$ or if $c_{3}(q) \geq c_{1}(q) \geq c_{2}(q)$, then $\lambda_{\rm max}=\lambda_3$ and $\lambda_{\rm min}=\lambda_1$ and so on. Moreover, in either case — whether the noise acts on one or both qubits — the ordering of the correlation components is preserved. That is, if $c_{i}>c_{j}$ initially, then $c_{i}(q)>c_{j}(q)$. This ensures that the identification of $\lambda_{\rm max}$ and $\lambda_{\rm min}$ is the same for both scenarios. Thus, the energy of passive state can be obtained as ${\rm Tr} \left[H \pi_{\rho_{\rm BDS}^{\rm q}} \right] = \lambda_{\rm min} - \lambda_{\rm max}=-\frac{1}{2}({\rm max} \{ c_{i}(q) \} + {\rm int} \{ c_{i}(q) \})$. So, the total ergotropy reads
\begin{align}
\mathcal W (\rho_{\rm BDS}^{\rm q}) &={\rm Tr} \left[H (\rho_{\rm BDS}^{\rm q} - \pi_{\rho_{\rm BDS}^{\rm q}}) \right]\\
&=\frac{1}{2}({\rm max} \{ c_{i}(q) \} + {\rm int} \{ c_{i}(q) \}),\label{total_erg_BDS_under_noise},
\end{align}
which is exactly the same as the average of geometrical quantum and classical correlations, given by Eq.~\eqref{GQC2}  and~\eqref{GCC2}, respectively so $\mathcal{W}(\rho_{\rm BDS}^{\rm q}) = \tfrac{1}{2} \left[ \mathcal{C}_{\rm G}(\rho_{\rm BDS}^{\rm q}) + \mathcal{Q}_{\rm G}(\rho_{\rm BDS}^{\rm q}) \right]$. $\blacksquare$

We emphasize that the proof is provided for the general form of the Bell-diagonal parameters $c_{i}(q)$, whose both qubits are affected by their local unital channels — whether the same or different — the eigenvalues of the state retain the form given in Eq.~\eqref{eig1_bell_diagonal}-~\eqref{eig4_bell_diagonal}. Consequently, the theorem remains valid in this more general scenario.

It can also be verified that when both qubits interact with a common noisy environment, the Bell-diagonal state remains invariant. For instance, in the case of a perfectly correlated bit-flip channel~\cite{sabale2024facets}, the corresponding Kraus operators can be written as $K_{0}^{\rm bf}=\sqrt{1-\frac{q}{2}} I \otimes I$ and $K_{1}^{\rm bf}=\sqrt{\frac{q}{2}} \sigma_{1} \otimes \sigma_{1}$. Then, it is straightforward to verify that this set of operators leaves the Bell-diagonal state Eq.~\eqref{Bell-diagonal-state} unchanged, thereby ensuring that the theorem remains valid under such collective noise as well. This invariance, however, does not generally hold for the other class of separable states with local coherence considered in this work. A comprehensive analysis of their behavior under common noise channels lies beyond the scope of the present manuscript and is left for future investigation.

\section{Frozen band of coherent work under bit flip}
\label{app_C}
In this section, we explain the origin of the white horizontal band in Fig.~\ref{contour_quantum_quantum}a, where the coherent ergotropy remains frozen under the bit-flip channel.

Since the local Hamiltonian has only two nonzero eigenenergies, $\pm 1$, according to  Eq.~\eqref{coerg_def} the coherent work reads
\begin{equation}
\begin{split}
\mathcal{W}^{\mathrm{C}}(q) = & {\rm Tr} \left[H (\pi_{\zeta(\rho)} - \pi_{\rho}) \right] =\\
&\min \{\lambda_{i}^{\rm d}(q) \} - \max \{\lambda_{i}^{\rm d}(q) \} - \min \{\lambda_{i}(q) \} + \max \{\lambda_{i}(q) \},\label{coh_erg_local}
\end{split}
\end{equation}
\\
where $\lambda_{i}(q)$ and $\lambda_{i}^{\rm d}(q)$ denote the eigenvalues of the noisy state $\rho(q)$ and its dephased counterpart, respectively. 

Without loss of generality, we focus on a single local state $\rho(a,c)$ where $c \in \Re$ f. The argument can be straightforwardly extended to the two-qubit separable state of Eq.~\eqref{quantum-quantum}. Under bit-flip noise, the eigenvalues are given by
\begin{equation}
\lambda_{1,2}(q) = \frac{1}{4}\!\left(1 \pm Z\right),\label{eigs_rho(q)}
\end{equation}
where $Z=\sqrt{\,4c^{2} + (-1+q)^{2} - 4a(-1+q)^{2} + 4a^{2}(-1+q)^{2}\,}$ and for the dephased state,
\begin{align}
\lambda^{d}_{1}(q) &= \frac{1}{4}\!\left[2 + 2a(-1+q) - q\right], \\
\lambda^{d}_{2}(q) &= \frac{1}{4}\!\left[-2a(-1+q) + q\right].
\end{align}
It is straightforward to verify that for $a = \frac{1}{2}$, the dephased state becomes degenerate, $\lambda^{d}_{1}(q) = \lambda^{d}_{2}(q)$. Consequently, the first two terms in Eq.~\eqref{coh_erg_local} cancel, and $\mathcal{W}^{\mathrm{\rm C}}(q) = \max \{\lambda_{i}(q) \} - \min \{\lambda_{i}(q)\}$. Moreover, when $a = \frac{1}{2}$, one finds 
$\min \{\lambda_{i}(q)\} = \frac{1}{4}(1 - 2c) = \min \{\lambda_{i}(0)\}$ and 
$\max \{\lambda_{i}(q)\} = \frac{1}{4}(1 + 2c) = \max \{\lambda_{i}(0)\}$, thus $\mathcal{W}^{\mathrm{\rm C}}(q) = \mathcal{W}^{\mathrm{\rm C}}(0)$, proving that the coherent ergotropy is frozen. $\blacksquare$

Similar band appear also under phase flip, as shown in Fig.~\ref{contour_quantum_quantum}c, which can be justified using the same logic as above.

\begin{figure*}[t]
\begin{center}
\includegraphics[width=\textwidth]{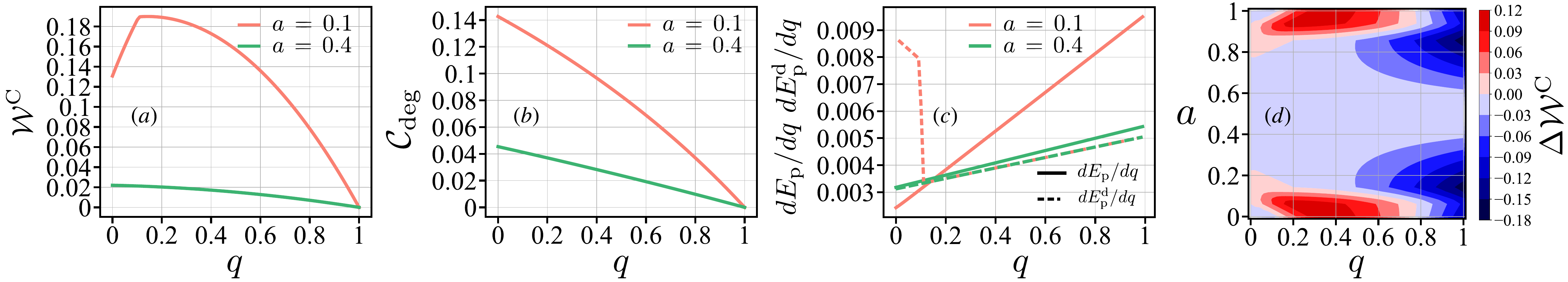}
\end{center}
\caption{(a) coherent ergotropy, (b) quantum coherence in degenerate subspace, (c) derivative of passive state energy $E_{\rm p}$ and $E_{\rm p}^{\rm d}$ and (d) difference in noisy and noiseless coherent ergotropy of the separable state Eq.~\eqref{quantum-quantum} and Hamiltonian~\eqref{interactive_Hamiltonian} under depolarizing channel. The other parameters are $c=0.3$, $d=0.2$.}
\label{depolarizing_appendix}
\end{figure*}

\section{Enhancement of coherent work under depolarizing channel}
\label{app_D}
In this Appendix we show that the coherent extractable work of the separable state in Eq.~\eqref{quantum-quantum} can be enhanced even under the depolarizing channel when the system Hamiltonian is chosen as
\begin{equation}
H = h(\sigma_x^{(1)} + \sigma_x^{(2)}) + J\, \sigma_x^{(1)} \otimes \sigma_x^{(2)} ,\label{interactive_Hamiltonian}
\end{equation}
whose eigenvalues are $-J, -J, -2h + J,$ and $2h + J$. The two eigenstates
\begin{equation}
|\psi^{-}\rangle = \tfrac{1}{\sqrt{2}}(|ge\rangle - |eg\rangle), \qquad
|\phi^{-}\rangle = \tfrac{1}{\sqrt{2}}(|gg\rangle - |ee\rangle),
\end{equation}
form a degenerate subspace.

To obtain an increase in coherent ergotropy, i.e., $\Delta \mathcal{W}^{\rm C} = \mathcal{W}^{\rm C}(q) - \mathcal{W}^{\rm C}(0) > 0$, 
the rate of variation of the passive states energy must satisfy
\begin{equation}
\frac{dE_{\rm p}}{dq} > \frac{dE_{{\rm p}}^{\rm d}}{dq},\label{dEp_dEpd}
\end{equation}
where $E_{\rm p}={\rm Tr}\left[H \pi_{\rho} \right] $ and $E_{\rm p}^{\rm d}={\rm Tr}\left[H \pi_{\zeta(\rho)} \right]$ denote the passive state energies of the total and dephased states, respectively.
Since $E_{\rm p}^{\rm d}$ depends solely on the populations in the energy basis, while $E_{\rm p}$ also depends on coherence, 
the presence of coherence---particularly within the degenerate subspace---can delay the spectral evolution of $\rho$, 
causing a lag in evolution of $E_{\rm p}$ with noise. When this lag is sufficiently strong, $\Delta \mathcal{W}^{\rm C}$ becomes positive, depending on noise strength $q$ and also population parameter $a$.
Fig.~\ref{depolarizing_appendix}a compares the coherent ergotropy for two population parameters, $a=0.1$ and $a=0.4$. 
For $a=0.1$, $\mathcal{W}^{\rm C}$ can grow within a finite range of depolarizing strength $q$, whereas for $a=0.4$ it decreases monotonically. 
Fig.~\ref{depolarizing_appendix}b shows coherence in the degenerate subspace, quantified as
\begin{equation}
\mathcal C_{\mathrm{\rm deg}} = 2\,\big|\langle\psi^{-}|\rho|\phi^{-}\rangle\big|,\label{coherence_degenerate}
\end{equation}
which is considerably larger in the case exhibiting enhanced coherent work.
The derivatives $dE_{\rm p}/dq$ and $dE_{{\rm p}}^{\rm d}/dq$ are also plotted in Fig.~\ref{depolarizing_appendix}c, in which 
for $a=0.1$, the condition $dE_{{\rm p}}^{\rm d}/dq > dE_{\rm p}/dq$ is fulfilled over the same noise interval where the coherent ergotropy is enhanced. 
Finally, Fig.~\ref{depolarizing_appendix}d presents a contour plot of $\Delta \mathcal{W}^{\rm C}$ versus $a$ and $q$, 
showing two regions at small and large $a$ where the coherent ergotropy is enhanced for most values of $q$, 
except at very small or very large values of $q$.

\begin{figure*}[t]
\begin{center}
\includegraphics[width=\textwidth]{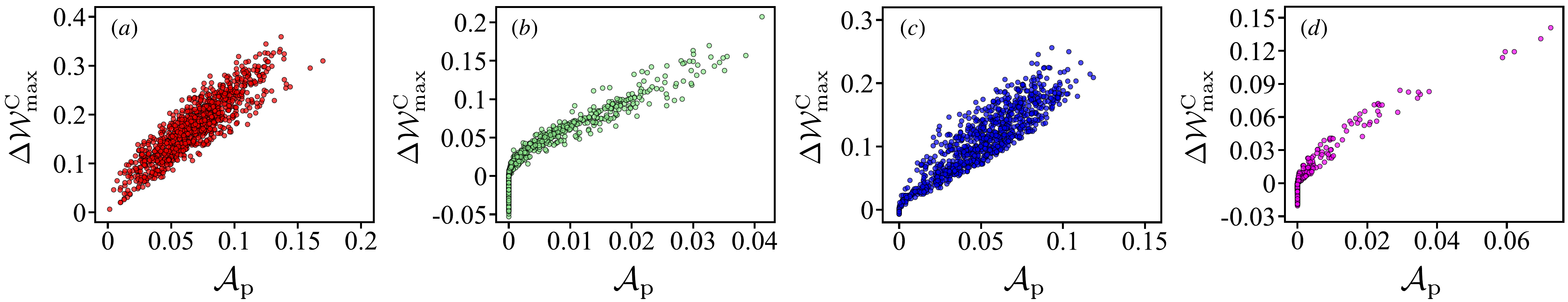}
\end{center}
\caption{ Maximum versus area of enhancement in coherent ergotropy for a randomized separable states Eq.~\eqref{general_two-qubit_separable} with $10^3$ number of samples for (a) bit flip, (b) amplitude damping, (c) phase flip and (d) depolarizing channel.}
\label{randomized_state}
\end{figure*}

\section{General two-qubit separable state with randomized parameters}
\label{app_E}
In this Appendix, we go beyond the separable state of Eq.~\eqref{quantum-quantum}, which was constructed as a permutation of two local states with identical populations and different coherences.  
Here we consider a general separable state of the form
\begin{equation}
\rho = \sum_i p_i \rho_i^{A} \otimes \rho_i^{B},\label{general_two-qubit_separable}
\end{equation}
where each local component is
\[
\rho_i^{j} = 
\begin{pmatrix}
a_i^j & c_i^j \\
c_i^j & 1 - a_i^j
\end{pmatrix},
\qquad j = A,B.
\]
The parameters $a_i^j$ and $c_i^j$ are randomized, independently sampled from uniform distributions, $a_i^j \in [0,1]$ and $c_i^j \in [0, c_{\max}]$, with $c_{\max} = \sqrt{a_i^j (1 - a_i^j)}$ ensuring physical validity.  
A total of $10^3$ random realizations are generated for each noise model.

To characterize the results, for each realization we study the maximum coherent-work enhancement over the noise strength $\Delta \mathcal{W}^{\rm C}_{\max}$ and the corresponding integrated enhancement area $\mathcal{A}_{\rm p}$ which quantifies the total range and magnitude of positive enhancement, respectively given by Eq.~\eqref{DeltaW_max} and~\eqref{enhancement_area}.

The results are shown in Fig.~\ref{randomized_state}. For the bit-flip channel (Fig.~\ref{randomized_state}a), all realizations exhibit an enhancement in both $\Delta \mathcal{W}^{\rm C}_{\max}$ and $A_p$, with a clear positive correlation between them.  
This observation aligns with Fig.~\ref{contour_quantum_quantum}a, in which large parameter region for bit-flip channel corresponds to positive increments in coherent ergotropy. A similar trend is observed for the phase-flip channel (Fig.~\ref{randomized_state}c), where over $95\%$ of the random states exhibit enhancement.  
In contrast, for the amplitude-damping and depolarizing channels (Fig.~\ref{randomized_state}b and~\ref{randomized_state}d, respectively), the number of realizations for which the coherent ergotropy grows, decreases—remaining above $57\%$ and $24\%$, respectively.  
The rest of samples form a vertical cluster along $A_p = 0$, indicating that for those instances the coherent ergotropy does not increase over the noise interval.
\section{Compatibility of the results with master equation formalism}
\label{app_F}
In this section, we verify that the results obtained using the discrete Kraus-operator description of Markovian channels are consistent with the continuous-time Lindblad master-equation approach (interaction picture) 
\begin{equation}
\frac{d\rho}{dt} = \sum_i \gamma_i \left( L_i \rho L_i^\dagger - \frac{1}{2} \{ L_i^\dagger L_i, \rho \} \right),
\label{master_dissipative}
\end{equation}
where $\gamma_i$ denotes the decay rate associated with each channel and $L_i$ are the corresponding Lindblad (collapse) operators that characterize the type of noise process acting on the system. As an illustrative example we derive the time dependence of the Bloch components for the bit-flip channel from the dissipative master equation and show explicitly how the result coincides with the discrete Kraus description once the usual relation between the continuous decay rate and the Kraus probability is enforced.

Consider a single qubit in generic Bloch representation, given by Eq.~\eqref{bloch-single-qubit}. Then the Lindblad generator for the bit-flip process is obtained by choosing $\sigma_x$ as the jump operator:
\begin{equation}
\dot{\rho}(t)=\gamma\big(\sigma_x\rho(t)\sigma_x-\tfrac{1}{2}\{\sigma_x^2,\rho(t)\}\big)
=\gamma\big(\sigma_x\rho(t)\sigma_x-\rho(t)\big).\label{bf_master}
\end{equation}
Solving Eq.~\eqref{bf_master} for Bloch vector components yields:
\begin{equation}
\begin{aligned}
n_1(t) &= n_1(0),\\
n_2(t) &= n_2(0)\,e^{-2\gamma t},\\
n_3(t) &= n_3(0)\,e^{-2\gamma t}.
\end{aligned}
\label{bf_bloch_time_dependent}
\end{equation}

\begin{figure*}[t]
\begin{center}
\includegraphics[width=\textwidth]{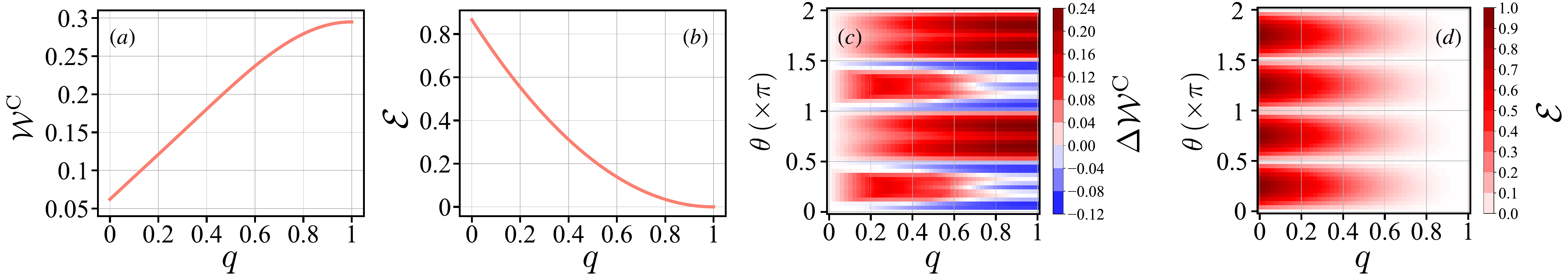}
\end{center}
\caption{(a) Coherent ergotrpy and (b) entanglement of $\ket{\tilde{\psi}(\theta)}$ (Eq.~\eqref{psi_tilde}) under bit flip channel with $\theta=2$. Contour plot of (c) coherent ergotropy enhancement and (d) entanglement versus the state parameter and noise strength. The Hamiltonin parameters are $h=0.5$ and $J=0.4$.}
\label{entangled_state}
\end{figure*}

On the other hand, Kraus representation of bit flip channel maps the Bloch vector as $\mathbf{n}\mapsto\mathbf{n}^{\rm bf}=(n_1,\,(1-q)n_2,\,(1-q)n_3)$ (see the main text). Thus, the two approaches coincide by enforcing
\[
e^{-2\gamma t}=1-q \;\Longrightarrow\; q(t)=1-e^{-2\gamma t}.
\]
Equivalently, for a given target Kraus probability (noise strength) $q$ and decay rate $\gamma$, the required physical time is
\begin{equation}
t = -\frac{1}{2\gamma}\ln(1-q),
\label{bf_time_from_q}
\end{equation}
for which the two formalisms coincide. Finally, substituting the time-dependent Bloch components Eq.~\eqref{bf_bloch_time_dependent} (or, equivalently, $q\mapsto q(t)=1-e^{-2\gamma t}$) into the previously obtained coherent ergotropy expression for the bit-flip channel (Eq.~\eqref{coerg_bit_flip_single}) results in the time-dependent form
\begin{equation}
\begin{split}
\mathcal W^{\rm C}(t) = &\frac{1}{2}\Big(\sqrt{\,n_1(0)^2 + n_2(0)^2 e^{-4\gamma t} + n_3(0)^2 e^{-4\gamma t}\,}\;-\\
&|n_3(0)|\,e^{-2\gamma t}\Big). \label{time_dependent_coerg_bit_flip}
\end{split}
\end{equation}
This expression is algebraically identical to evaluating the Kraus operator based formula at $q=q(t)$ up to proper physical time  therefore the Kraus and master equation descriptions give the same when evaluated at the proper physical time given by Eq.~\eqref{bf_time_from_q}. Therefore, both result in the same frozen and enhancement criteria for coherent ergotropy.

As a second example, we consider the amplitude-damping channel, whose jump operator is the lowering operator $\sigma^{-}$. The corresponding time evolved Bloch parameters are obtained as
\begin{equation}
\begin{aligned}
n_{1}(t)&=n_{1}(0)e^{-\gamma t/2},\\
n_{2}(t)&=n_{2}(0)e^{-\gamma t/2},\\
n_{3}(t)&=n_{3}(0)e^{-\gamma t}-\big(1-e^{-\gamma t}\big).
\end{aligned}
\label{eq:ad_bloch_time}
\end{equation}
Similar correspondence between Kraus operator and master equation approach is then defined 
\[
1-q = e^{-\gamma t}
\quad\Longleftrightarrow\quad
q(t)=1-e^{-\gamma t}.
\]
For the other noise channels, a similar correspondence can be obtained by selecting the relevant jump operators in the master equation. This analyze can be simply extended to the case of multipartite systems. 

\section{Noise-assisted enhancement of coherent ergotropy for entangled states: an example}
\label{app_G}
So far, our analysis of two- and multi-qubit systems has been restricted to separable states. This choice was motivated by the central aim of the present work, namely to demonstrate that noise-assisted enhancement of coherent ergotropy does not rely on entanglement, but can already occur for separable states provided that sufficient local coherence is present. In particular, we showed that separable states equipped with local coherence can exhibit an increase of coherent ergotropy under various Markovian noise channels, including flip-type and depolarizing noise, depending on the system Hamiltonian.

Nevertheless, it is natural to ask whether similar noise-assisted enhancement can also arise when the initial state is entangled. In this Appendix, we provide a concrete numerical example demonstrating that entanglement does not preclude the enhancement of coherent ergotropy, while emphasizing that local coherence remains a necessary ingredient. To this end, we consider an entangled two-qubit pure state of the form
\begin{equation}
\ket{\psi(\theta)} = \cos\theta\,\ket{\rm gg} + \sin\theta\,\ket{\rm ee},
\end{equation}
which is entangled for $\theta \neq n \pi/2$ where $n \in \mathbb{Z}$. Now we let both qubits undergo local Hadamard operations and define the transformed state
\begin{equation}
\ket{\tilde{\psi}(\theta)} = (U_H \otimes U_H)\ket{\psi(\theta)},\label{psi_tilde}
\end{equation}
where $U_H=1/\sqrt{2} \big( \ket{\rm g}\bra{\rm g} + \ket{\rm g}\bra{\rm e} + \ket{\rm e}\bra{\rm g} - \ket{\rm e}\bra{\rm e} \big)$. We find that $\ket{\tilde{\psi}(\theta)}$ can exhibit noise-assisted enhancement of the coherent ergotropy with respect to the Hamiltonian
\begin{equation}
H = h(\sigma_z^{(1)} + \sigma_z^{(2)}) + J\, \sigma_x^{(1)} \otimes \sigma_x^{(2)}.\label{Hamiltonian_for_entangled}
\end{equation}
For simplicity, we restrict the discussion here to the bit-flip channel, although similar behavior can be obtained for other noise models. Fig.~\ref{entangled_state}a and~\ref{entangled_state}b show the behavior of the coherent ergotropy and the entanglement, quantified by the quantum concurrence~\cite{wootters1998entanglement,hill1997entanglement}, respectively, as functions of the bit flip noise strength $q$. For the representative value $\theta=2$, the coherent ergotropy increases monotonically with increasing noise strength and reaches its maximum at the maximal noise limit $q=1$. In contrast, the concurrence decreases monotonically under the action of the bit flip channel and vanishes at $q=1$ which is consistent with the dissipative nature of the channel. Despite these opposite trends, there exists an intermediate range of noise strengths for which the state remains entangled while the coherent ergotropy is simultaneously enhanced. This observation demonstrates that noise-assisted enhancement of coherent ergotropy is not limited to separable states and can also occur when the system is entangled. To further elucidate the interplay between entanglement and coherent ergotropy, Figs.~\ref{entangled_state}c and~\ref{entangled_state}d show contour plots of the coherent ergotropy and the concurrence, respectively, as functions of the state parameter $\theta$ and the bit flip noise strength $q$. As mentioned earlier, the state becomes separable if $\theta=n\pi/2$ when $n\in \mathbb{Z}$, and this feature is clearly reflected in both panels. In particular, along these values of $\theta$, the coherent ergotropy does not exhibit any enhancement under the bit flip channel. For $\theta \neq \pi/2$, regions appear where the state is entangled and, at the same time, the coherent ergotropy increases with the noise strength. Moreover, both quantities display a periodic dependence on $\theta$, inherited from the structure of the initial state. However, their behavior as functions of the noise strength is qualitatively different. While entanglement is progressively suppressed by the bit flip channel and always vanishes in the strong noise limit $q \to \infty$, the coherent ergotropy can remain nonzero in this regime and may even reach its maximum value, depending on $\theta$. These observations highlight that entanglement and coherent ergotropy respond differently to environmental noise. Although regions exist where entanglement and increasing coherent ergotropy coexist, the maximal enhancement of coherent ergotropy occurs in regimes where entanglement has already been completely destroyed. This clearly suggests that while entanglement may not be the primary resource driving the noise-assisted enhancement of coherent ergotropy, the presence of entanglement does not preclude such enhancement. Instead, the behavior of coherent ergotropy is mainly determined by the structure of coherence in the energy eigenbasis, which can be favorably modified by noise irrespective of whether the state is separable or entangled.

\end{document}